%% file: paper.tex
\documentclass[usenatbib,fleqn]{mnras}
\usepackage{graphicx,hyperref,amsmath,amssymb,xspace}
\usepackage[dvipsnames]{xcolor}

\title[Internal kinematics of globular clusters from Gaia]
{Systematic errors in \textit{Gaia} DR2 astrometry and their impact on measurements of internal kinematics of star clusters}
\author[E. Vasiliev]{Eugene Vasiliev$^{1,2}$\thanks{E-mail: eugvas@lpi.ru}\\
$^1$Institute of Astronomy, Madingley road, Cambridge, CB3 0HA, UK\\
$^2$Lebedev Physical Institute, Leninsky prospekt 53, Moscow, 119991, Russia}

\newcommand{\Gaia}{\textit{Gaia}\xspace}
\newcommand{\boldmu}{\boldsymbol\mu}
\newcommand{\diag}{\mathop{\mathrm{diag}}}
\newcommand{\kms}{km\:s$^{-1}$\xspace}
\newcommand{\masyr}{mas\:yr$^{-1}$\xspace}
\DeclareMathOperator{\sinc}{sinc}

\begin{document}
\maketitle

\begin{abstract}
We use stellar proper motions (PM) from \Gaia Data Release 2 for studying the internal kinematics of Milky Way globular clusters. 
In addition to statistical measurement errors, there are significant spatially correlated systematic errors, which cannot be ignored when studying the internal kinematics.
We develop a mathematically consistent procedure for incorporating the spatial correlations in any model-fitting approach, and use it to determine rotation and velocity dispersion profiles of a few dozen clusters. We confirm detection of rotation in the sky plane for $\sim10$ clusters reported in previous studies, and discover a few more clusters with rotation amplitudes exceeding $\sim 0.05$~\masyr. However, in more than half of these cases the significance of this rotation signature is rather low when taking into account the systematic errors. We find that the PM dispersion is not sensitive to systematic errors in PM, however, it is quite sensitive to the selection criteria on the input sample, most importantly, in crowded central regions. When using the cleanest possible samples, PM dispersion can be reliably measured down to 0.1~\masyr for $\sim60$ clusters.
\end{abstract}


\section{Introduction}

The second data release (DR2) of the \Gaia mission \citep{Brown2018} expanded the field of astrometry with an enormous dataset containing more than a billion stars with measured parallaxes and proper motions (PM). For stars brighter than magnitude $G=16$, the statistical uncertainty in PM is at the level 0.1~\masyr, corresponding to an error in velocity of $0.5\,(D/1\mbox{ kpc})$~\kms, where $D$ is the distance to the star. A typical globular cluster at a distance 10~kpc has many stars with magnitudes $G\le16$ on the upper part of the giant branch or on the horizontal branch. The measurement uncertainties in individial velocities of $\sim5$~\kms are comparable to the internal velocity dispersion, but by averaging over many stars, one could hope to determine the intrinsic velocity dispersion and the mean velocity with a precision of $1-2$~\kms \citep{Pancino2017}, comparable to that of studies based on line-of-sight velocities.

Unfortunately, the averaging will not help to reduce systematic errors, which appear to be present in the \Gaia astrometry. They are most easily illustrated by analyzing the measured values of parallax and PM of distant quasars or stars in the Large Magellanic cloud (LMC), as shown on Figures~12, 13 in \citet{Lindegren2018} or Figures~16, 17 in \citet{Helmi2018}. Even after averaging over many sources in the same area of the sky, there remain residuals at the level of a few$\times10^{-2}$~mas or \masyr, visible both as large scale variations on the sky plane, and as peculiar patterns at scales $0.5-1^\circ$, associated with the \Gaia scanning law. Attempting to compensate these residual errors in ordinary stars is very challenging \citep[Section~4]{Arenou2018}, as the errors may depend on various other parameters such as magnitude and colour, and the density of extragalactic sources is not high enough to evaluate these variations in any particular region on the sky. Nevertheless, \citet{Chen2018} managed to estimate and compensate the systematic offset in parallax for two globular clusters, NGC~104 (47~Tuc) and NGC~362, which lie close to the Small Magellanic cloud, using a two-step correction procedure that involved both scarse distant quasars and more numerous stars of the background galaxy. This correction is hardly possible in other cases, and hence the systematic errors should be treated as random variables drawn from some global distribution, and added to the statistical errors.

It is particularly important to take into account these spatially correlated systematic errors in PM when considering the internal kinematics of star clusters (variation of stellar PM due to internal rotation and velocity dispersion), since the systematic offsets are not constant for the entire cluster, but rather vary across its area. Recently, several studies measured the internal kinematics of globular clusters from \Gaia DR2 PM: \citet{Bianchini2018} detected rotation signatures in 11 clusters (some of them were also mentioned in \citealt{Helmi2018}), \citet{Baumgardt2019} presented a comprehensive collection of PM dispersion profiles for more than 100 clusters, \citet{Jindal2019} measured rotation, dispersion and anisotropy profiles for 10 clusters, and \citet{Sollima2019} investigated 3d rotation patterns in 60 clusters, combining \Gaia PM with line-of-sight velocity measurements. In all cases, though, the problem of systematic errors was not addressed in detail. In the present study, we attempt to fill this gap.

The paper has two main objectives. 
We develop a mathematically coherent formalism for incorporating the systematic errors in the analysis of internal kinematics of star clusters in Section~\ref{sec:syserr}, with more specific mathematical details deferred to the \hyperref[sec:appendix]{Appendix}. We then apply this approach to the problem of measuring the profiles of mean PM (including internal rotation) and its dispersion in Milky Way globular clusters in Section~\ref{sec:globclus}. We perform extensive tests on both observed and simulated data, and conclude that the rotation could be reliably measured even in the presence of systematic errors, if its peak amplitude exceeds $\sim0.05$~\masyr, while the internal dispersion is less sensitive to the systematic errors (although it may be biased by other factors) and could be measured down to $\sim 0.1$~\masyr.
In this study, we focus on globular clusters, but the approach should be applicable to any other stellar system (open clusters or dwarf galaxies, although for the latter, the precision of current \Gaia data is insufficient to draw any conclusions about their internal kinematics).

\section{Systematic errors in astrometry}  \label{sec:syserr}

\begin{figure}
\includegraphics{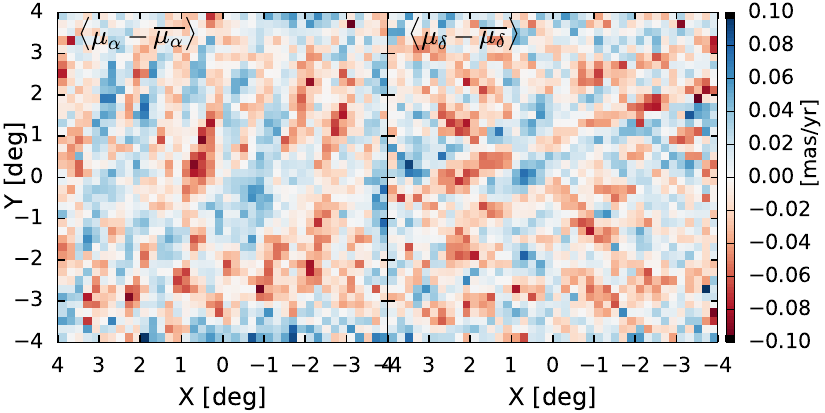}
\caption{Residuals in the mean PM of stars in the $8\times8^\circ$ field centered on the LMC. Shown are the average values of $\mu_\alpha$ (left) and $\mu_\delta$ (right) in $0.2\times0.2^\circ$ bins, after subtracting the mean PM trend represented by a bicubic function fitted to all stars in the field. Some of the residuals may be caused by the real kinematic features on smaller scales than the mean trend (such as the influence of the bar), but most of the remaining variation is likely due to systematic errors, tracing the scanning pattern. (See also similar illustrations in Figure~17 of \citealt{Helmi2018} and Figure~13 of \citealt{Lindegren2018}).
} \label{fig:pm_residuals_lmc}
\end{figure}

The measurements of astrometric parameters (parallax $\varpi$ and two components of proper motion $\mu_\alpha \equiv \mathrm{d}\alpha/\mathrm{d}t\;\cos\delta$, $\mu_\delta \equiv \mathrm{d}\delta/\mathrm{d}t$) in the \Gaia data carry a statistical error (quoted in the catalogue as a full covariance matrix between the three parameters) and a systematic error that is not easily accessible. By definition, the systematic error cannot be eliminated by averaging over many sources that are drawn from the same population (i.e., can be assumed to have the same true values of $\varpi$ and $\boldsymbol\mu$), and hence it becomes dominant for large enough sample sizes. Moreover, this error depends on the position and is correlated in nearby locations on the sky in a way that is determined by the \Gaia scanning pattern.

An estimate of the magnitude of the systematic error and its covariance between two different points on the sky can be obtained by analyzing the measurements of an ensemble of sources with known true values. The $\sim 0.5$ million quasars detected by \Gaia across nearly the entire sky (excluding the Galactic disc) are a prime choice for such a sample, since their true parallaxes and PM are zero. 
Another example is the LMC, which has a high density of stars situated at similar distances (true parallax $\varpi \simeq 0.02$~mas is much smaller than the measurement errors, and the intrinsic spread in parallax is $\sim10\times$ smaller). Of course, these stars have a nonzero mean PM, which furthermore varies across the surface of the galaxy, but after subtracting the mean trend, one can clearly see the residual variations on scales $\sim 0.5-1^\circ$ (Figure~\ref{fig:pm_residuals_lmc}).

The covariance function is, in general, a function of two positions on the sky, and possibly the magnitude and colour of the sources, which makes it extremely challenging both to estimate and to apply. One could nevertheless make progress by assuming that it depends only on the angular separation $\theta$ between the two sources: $V(\theta_{ij}) = \big\langle (\mu_i - \mu_i^\mathrm{true})  (\mu_j - \mu_j^\mathrm{true}) \big\rangle$. The obvious directional dependence of residuals plotted in Figure~\ref{fig:pm_residuals_lmc} demonstrates that this assumption does not convey the full complexity of spatial correlations, but it already allows one to qualitatively estimate the systematic errors. We also assume that $V$ is the same for both PM components and is zero for the mixed covariance $(\mu_{i,\alpha} - \mu_{i,\alpha}^\mathrm{true}) (\mu_{j,\delta} - \mu_{i,\delta}^\mathrm{true})$ -- this follows from the symmetry considerations, namely the invariance w.r.t. rotation of the coordinate frame.

The average density of only a dozen quasars per square degree is barely sufficient to robustly measure the covariance function on small scales. The typical statistical uncertainty on individual PM measurements is $\sim 0.5-1$~\masyr at quasar magnitudes, and the number of quasar pairs over the entire sky with angular separation less than $\theta$ is $\sim1.2\times10^7\: (\theta/1^\circ)^2$, resulting in the uncertainty on $V(\theta)$ at a level $0.0002\: (\theta/1^\circ)^{-1} \mbox{ [mas\:yr}^{-1}]^2$.
\citet{Lindegren2018}, Section 5.4, plot the PM covariance function $V(\theta)$ averaged over $0.125^\circ$-wide bins in $\theta$ (their Figure~15), and provide an estimate of the mean trend at scales $\theta\gtrsim 0.5^\circ$.
Figure~\ref{fig:covfnc} plots the same data for quasars in green, and the average pairwise residuals in PM of LMC stars in yellow. The latter show a smaller-amplitude covariance, which nearly vanishes for $\theta\gtrsim 0.5^\circ$ due to the subtraction of the mean PM before computing pairwise correlations. Nevertheless, the overall oscillatory behaviour and a strong increase of $V(\theta)$ at small separations is clear from both datasets. 

A mathematically consistent description of pairwise correlations imposes certain restrictions on the form of the function $V(\theta)$. If the spatial variable $\theta$ were a real axis $(-\infty,\infty)$, then $V(\theta)$ must have a non-negative Fourier transform in order to be a valid covariance function. When $\theta$ is the angular separation on a 2d sphere, the analogous requirement is non-negativity of the Legendre integral transform%
\footnote{Not to be confused with an unrelated concept of Legendre transformation. Legendre transform is essentially the spherical-harmonic transform of a function that is independent of $\phi$, hence it contains only $m=0$ terms for any $l$.}%
: $C_l \equiv \int_{-1}^1 \mathrm{d}\cos\theta\;P_l(\cos\theta)\;V(\theta) \ge 0$, where $P_l(x)$ are the Legendre polynomials. For instance, $V(\theta) = \exp(-\theta/\theta_0)$ or $V(\theta) = \exp\big(-[\theta/\theta_0]^2\big)$ are valid choices, whereas $V(\theta)=\cos(k\theta)$ is only valid when $k$ is an integer. A piecewise-constant function (such as the value of $V$ in the nearest bin estimated from the quasar sample) is also not a valid choice. This motivates the need for a suitable analytic approximation of the covariance function that would satisfy the condition of non-negativity of its Legendre transform. Two possible choices are shown in Figure~\ref{fig:covfnc} as blue and red lines, and have the following simple analytic forms (the numerical value is measured in units of $[\mbox{mas\:yr}^{-1}]^2$):

\begin{figure}
\includegraphics{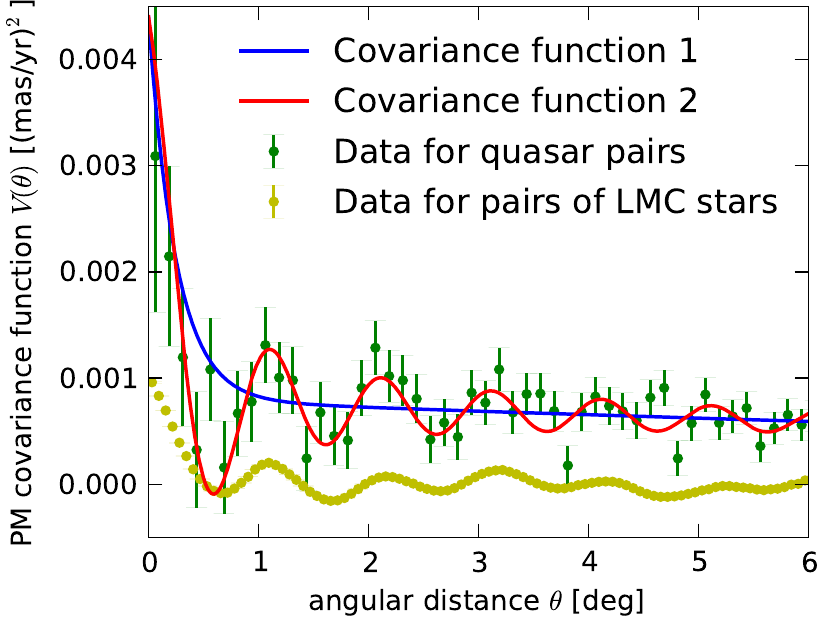}\vspace*{-2mm}
\caption{Covariance function of systematic errors in PM as a function of angular separation between pairs of sources. \protect\\
Green points with error bars show the covariances between PM of quasars, mostly in the range of magnitudes $18\le G \le 20$, averaged over bins of width $0.125^\circ$. The typical systematic error is $[0.03\mbox{ mas\:yr}^{-1}]^2$ on angular scales $\gtrsim0.5^\circ$, and roughly twice larger at small scales. Yellow points plot the same data for the LMC stars shown on Figure~\ref{fig:pm_residuals_lmc} (error bars are smaller than the marker size), which have magnitudes $16\le G \le 18$. Since the mean trend was subtracted for these stars, their covariance on scales $\gtrsim0.5^\circ$ is close to zero, and is smaller than the quasar sample in the limit of small separation, but follows a similar overall oscillatory trend. Blue and red curves are the two analytic approximations given by Equation~\ref{eq:covfnc}.
} \label{fig:covfnc}
\end{figure}

\begin{equation}  \label{eq:covfnc}
\begin{aligned}
V_1(\theta) &= 0.0008 \exp(-\theta/20^\circ) + 0.0036\exp(-\theta/0.25^\circ) , \\
V_2(\theta) &= 0.0008 \exp(-\theta/20^\circ) + 0.004\sinc(\theta/0.5^\circ + 0.25) .\!\!\!\!\!
\end{aligned}
\end{equation}

The first term in each function is the same as suggested by \citet{Lindegren2018} for the large-scale errors, while the second term captures the increase of correlated errors at small scales and in the limit $\theta=0$ corresponds to a systematic error in $\mu$ of 0.066~\masyr, again in accordance with the above paper.
The first choice of covariance function, $V_1$, is non-oscillatory and somewhat more conservative, as illustrated by the tests described below, while the second one, $V_2$, more closely resembles the actual quasar data with its degree-scale oscillatory patterns. 

The spatially correlated systematic errors have an obvious impact on the determination of the mean PM or its internal variation for any sample of stars, for instance, a star cluster or a satellite galaxy. A simple approach is to take the average value of $V(\theta_{ij})$ for all pairs of sources $i,j$ in the sample as the estimate of the systematic uncertainty of the mean PM%
\footnote{This is the recipe suggested in the presentation on the DR2 astrometry by L.Lindegren et al., available at \url{https://www.cosmos.esa.int/web/gaia/dr2-known-issues}.}.
However, this is not optimal in the sense that it uses an unweighted average, while the standard approach for the statistical errors is to weigh each star's contribution to the average PM by the inverse square of its measurement uncertainty $\epsilon$, giving more weight to stars with smaller errors:
\begin{equation*}
\langle \mu \rangle = \frac{\sum_i \mu_i / \epsilon_{\mu,i}^2}{\sum_i 1 / \epsilon_{\mu,i}^2} .
\end{equation*}
The corresponding weighting in the case of systematic errors does not have such a simple form, and involves the inversion of the overall $N\times N$ covariance matrix for all pairs of stars, as explained in the \hyperref[sec:appendix]{Appendix}. Figure~\ref{fig:mean_pm_sys_err} illustrates the systematic uncertainty on the mean PM for a cluster of angular size $\theta$ (with a Gaussian surface density profile), for the two choices of $V(\theta)$. For compact stellar systems with the radius less than a few arcminutes, the uncertainty in $\langle\mu\rangle$ is essentially $\sqrt{V(0)}\simeq 0.066$~\masyr, while for systems spanning more than half a degree it is roughly twice smaller. Weighted averaging produces somewhat lower uncertainties, and the oscillatory covariance function $V_2(\theta)$ results in a faster transition from the point-like to the finite-radius regimes. Hence we consider the non-oscillatory function $V_1$ to be a more conservative choice, and use it in the rest of the paper.

For completeness, we also derived analytic approximations for the spatial covariance function of systematic parallax errors (after subtracting the mean parallax offset of $-0.029$~mas, as suggested by \citealt{Lindegren2018}). These are described by the same functional forms as Equation~\ref{eq:covfnc}, with the first term replaced by $0.0003\exp(-\theta/14^\circ)$, and the second terms being twice smaller.

\begin{figure}
\includegraphics{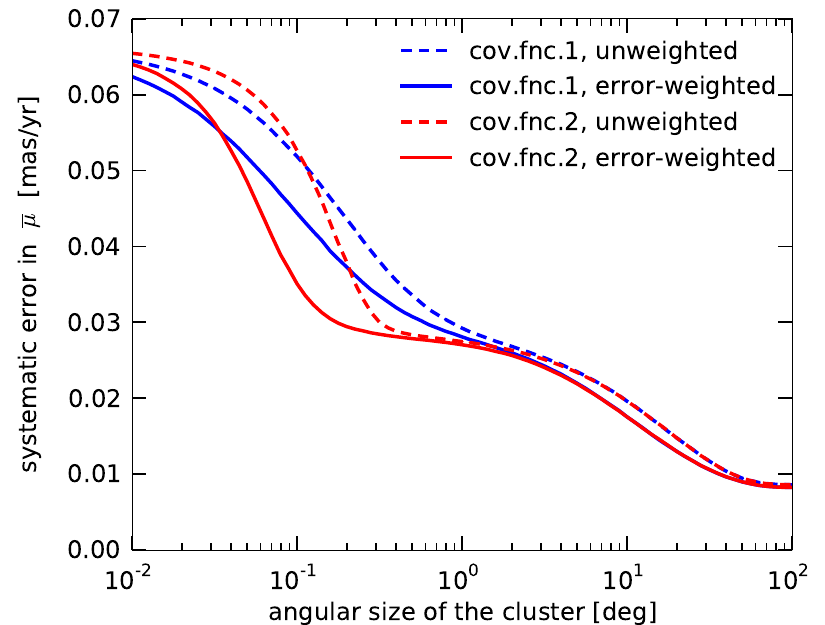}
\includegraphics{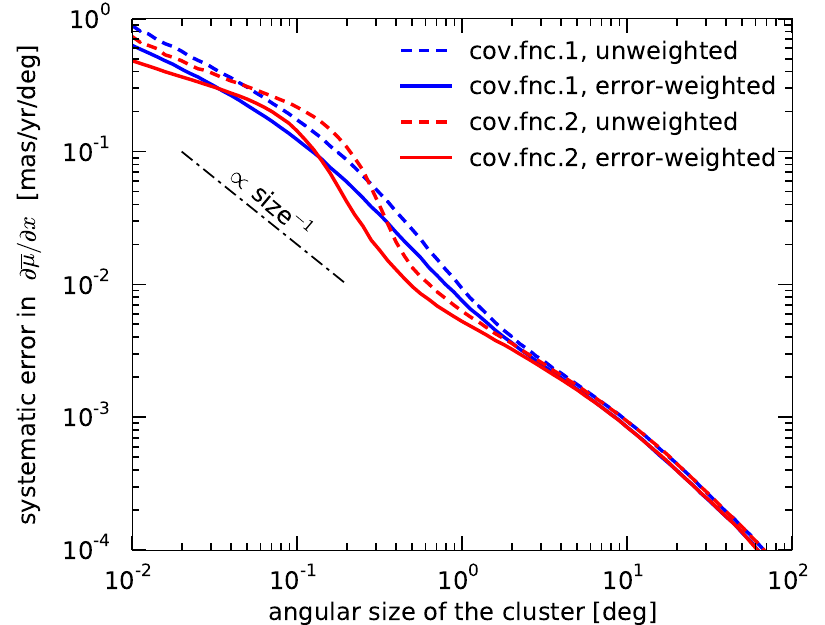}
\caption{Systematic uncertainty of the mean PM (top panel) and its spatial gradient (bottom panel) for a cluster of the given angular size $s$ and a Gaussian surface density profile. 
Blue and red curves show two choices of covariance function in Equation~\ref{eq:covfnc} ($V_1$ and $V_2$, correspondingly); dashed lines show unweighted estimates and solid lines -- error-weighted estimates as described in the Appendix (considering only the systematic errors but neither statistical errors nor the intrinsic PM dispersion). For small angular sizes, the uncertainty in the mean PM approaches the limiting value 0.066~\masyr, and the uncertainty in the PM gradient is $\simeq 0.01\,s^{-1}$~\masyr for all cluster sizes.
} \label{fig:mean_pm_sys_err}
\end{figure}

\section{Analysis of internal kinematics of globular clusters}  \label{sec:globclus}

We now consider the problem of determining the internal kinematics (mean internal PM and its dispersion) of globular clusters from the \Gaia data. Since the internal PM gradients (such as rotation signature) are produced by differences in the mean PM between nearby regions of the sky, they are quite sensitive to the spatially correlated systematic errors. One of the goals of this paper is to test how well they could be measured, and how much bias can one expect when ignoring the systematic errors.

\subsection{Membership determination}  \label{sec:membership}

As a first step, we obtain the list of cluster members. We use the following selection criteria for the initial list of stars:
\begin{itemize}
\item Parallax $\varpi$ consistent with the expected value for the given distance $D$ to the cluster within 3 times the statistical uncertainty $\epsilon_\varpi$ quoted in the \Gaia catalogue: $\varpi < 1/D + 3\epsilon_\varpi$ (adding a cut at $\varpi > 1/D - 3\epsilon_\varpi$ makes a negligible difference).
\item High quality of astrometric solution: \texttt{astrometric\_excess\_noise}${}<1$, and renormalized unit weight error ${}<1.2$ (see \citealt{Lindegren2018TN} for the definition of this parameter). This should eliminate most binary stars, as there is no special treatment for their internal motion in DR2, and they appear to be poorly fit by a straight-line motion model, resulting in large residual errors. \citet{Bianchini2016} found the effect of unresolved binaries on the PM dispersion to be relatively small.
\item No significant crowding that mainly affects faint stars in dense central regions of clusters: \texttt{phot\_bp\_rp\_excess\_factor}${}<1.3 + 0.06\;\mbox{\texttt{bp\_rp}}^2$, following the recommendations given in Appendix C of \citet{Lindegren2018}. 
\end{itemize}
We find that the last cut most strongly affects the central regions of dense clusters, often removing more than a half of stars, but without it, the PM dispersion profiles are significantly biased up, indicating that the statistical errors are likely underestimated for stars that are affected by crowding.

After selecting all stars satisfying the above criteria in the field of each cluster, we use the two-component Gaussian mixture approach from \citet{Vasiliev2019} to determine the membership probabilistically from their clustering on the sky and in the PM space, as described in Appendix~\ref{sec:membership_details}.

Out of 150 clusters in the catalogue, around a half contain enough stars ($N\gtrsim 100$) and are sufficiently close that the PM dispersion is above 0.1~\masyr, which we take as the minimum value that could be reliably determined.
For a few richest clusters -- NGC 104 (47 Tuc), NGC 5139 ($\omega$ Cen), NGC 6397 and NGC 6752 -- we limit the number of considered stars to $N=10^4$, picking up the brightest ones which have smaller PM uncertainties and hence are most informative (this is motivated by a steep scaling of the computational cost as $N^3$). Since the masses of giant stars are very similar across all magnitudes, we don't expect any magnitude dependence in their kinematics.

\subsection{The model}  \label{sec:model}

The model for internal kinematics is intentionally not based on any dynamical considerations and is purely data-driven. In essense, we estimate the mean PM and its dispersion in several intervals of projected radii, but instead of binning the data points, we represent the PM profiles in terms of smoothly varying functions (cubic splines) defined by their values at grid points. This reduces the dependence of the result on the choice of bins and more appropriately takes into account the spatially correlated systematic errors across the entire dataset. 

For each cluster, we convert the positions and PM of stars into a reference frame with origin at the cluster centre (taken from the catalogue; the variation between centre coordinates determined in different studies is typically at the level of a few arcseconds, which has negligible impact on the measured profiles at larger radii). The coordinates on the sky plane in this transformed system are denoted as $\boldsymbol{x}\equiv\{x,y\}$ (measured in units of angle), and the PM as $\{\mu_x,\mu_y\}$; the transformation is given, e.g., by Equation~2 in \citet{Helmi2018}.
Geometric sizes of clusters are small enough that one may neglect second-order corrections from the curvilinear coordinate transformation and split the PM of each star into the centre-of-mass PM of the entire cluster, and the internal PM drawn from some model distribution.
We assume that this internal PM distribution depends only on the projected distance $R\equiv\sqrt{x^2+y^2}$ from the cluster centre, and is described by a Gaussian with a mean value $\boldmu^\mathrm{mod}(\boldsymbol x)$ and an isotropic dispersion $\sigma^\mathrm{mod}_\mu(R)$. The mean PM consists of two components: radial $\mu_R \equiv (x\mu_x+y\mu_y)/R$ and tangential (rotational) $\mu_t \equiv (y\mu_x-x\mu_y)/R$. The former is positive when the cluster expands radially, and the latter is positive when the cluster rotates counter-clockwise on the sky.

The mean PM in the radial direction is expected to be caused by the perspective expansion or contraction, resulting from the motion of the cluster along the line of sight: \vspace*{-3mm}
\begin{equation}  \label{eq:mu_radial}
\frac{\mu_R^\mathrm{mod}}{\mbox{mas\:yr}^{-1}} =
-6.14\times10^{-5} \frac{R}{\mbox{arcmin}} \; \frac{v_\mathrm{los}}{\mbox{km\:s}^{-1}} \;
\left[\frac{D}{\mbox{kpc}}\right]^{-1},
\end{equation}
where $D$ is the distance to the cluster, and $v_\mathrm{los}$ is the line-of-sight velocity (positive if the cluster recedes). In fitting the data, we assume a linear dependence of $\mu_R^\mathrm{mod}$ on $R$, but allow $v_\mathrm{los}$ to be a free parameter, subsequently comparing it with the true line-of-sight velocity known from spectroscopic measurements.
In converting $v_\mathrm{los}$ to PM, we use the distances taken from the catalogue; their uncertainties at the level of few per cent are typically much smaller than the derived uncertainties on the PM profiles.

The mean PM in the tangential direction $\mu_t^\mathrm{mod}$ is assumed to depend only on $R$, but we consider a flexible form for it, represented by a cubic spline with $N_R$ nodes. We choose $N_R$ depending on the number of stars $N$ in our dataset: 3 if $N<300$, 4 if $N<1000$ and 5 otherwise. The first node is always at $R=0$, and we place the remaining nodes at radii enclosing a fixed percentage of stars (e.g., 12\%, 50\% and 95\% for $N_R=4$), aiming at a good balance between spatial coverage and resolution.
Since we use smooth spline profiles and evaluate the likelihood of measuring the PM of each star at its true distance from the cluster centre, not rounded to the nearest grid node, the results are much less sensitive to the particular choice of grid nodes, unlike the conventional binning approach.

The PM dispersion $\sigma^\mathrm{mod}_\mu(R)$ is represented by a similar spline defined on the same grid in $R$. The values of $\sigma_\mu$ at grid nodes are required to be positive (but not necessarily monotonic with radius), and the values of $\mu_t^\mathrm{mod}$ are not constrained in any way. We extrapolate both $\mu_t$ and $\sigma_\mu$ as constant beyond the last grid node. 

In total, the model is described by the following parameters: $N_R$ values of $\sigma^\mathrm{mod}_\mu$ at all grid nodes, $N_R-1$ values of $\mu^\mathrm{mod}_t$ at all nodes except the origin (where it is set to zero), and three components of the cluster centre-of-mass motion: $\overline\mu_x, \overline\mu_y$ and $v_\mathrm{los}$ (the latter only affects the radial PM component, since we do not use the line-of-sight velocities in the fit).

The optimal values of these parameters are found by maximizing the log-likelihood of the model, which is given by the following expression (up to a constant factor):
\begin{equation}
\ln\mathcal L = -{\textstyle\frac12} \ln\det\mathsf\Sigma - {\textstyle\frac12}
(\boldmu - \boldmu^\mathrm{mod})^T\; \mathsf\Sigma^{-1} \; (\boldmu - \boldmu^\mathrm{mod}) .
\end{equation}
Here the vectors $\boldmu$ and $\boldmu^\mathrm{mod}$ of length $2N$ contain the measured values of PM of all $N$ stars and the mean PM predicted by the model at each star's position, correspondingly.
$\mathsf\Sigma$ is the $2N\times 2N$ covariance matrix which combines the statistical errors for each star, correlated systematic errors for all pairs of stars, and the intrinsic PM dispersion $\sigma_\mu^\mathrm{mod}$. We use the Markov Chain Monte Carlo (MCMC) code \textsc{emcee} \citep{ForemanMackey2013} to explore the parameter space and determine confidence intervals. The mathematical details of the model-fitting approach are described in the Appendix~\ref{sec:fitting_details}.

In addition, we also considered series of models with $\mu_t^\mathrm{mod}(R)$ and/or $\sigma_\mu^\mathrm{mod}$ represented in a parametric form rather than a spline. For the former, we use Equation~1 in \citet{Bianchini2018}:
\begin{equation}  \label{eq:meanmu_parametric}
\mu_t^\mathrm{mod}(R) = \mu_t^\mathrm{peak} \;
\frac{2\; R/R_\mathrm{peak}}{1 + (R/R_\mathrm{peak})^2} ,
\end{equation}
and for the latter, the velocity dispersion profile of a Plummer sphere:
\begin{equation}  \label{eq:sigma_parametric}
\sigma_\mu = \frac{\sigma_\mu^\mathrm{centre}}{\big[1 + (R/R_\sigma)^2 \big]^{1/4}} .
\end{equation}

The resulting best-fit profiles generally agreed well with the free-form ones, but had somewhat narrower confidence intervals, hence we opted not to present them. The best-fit parameters for each cluster, derived from these fits, were used to generate the mock data, as explained in the following section.

\subsection{Tests on mock datasets}  \label{sec:mock}

\begin{figure}
\includegraphics{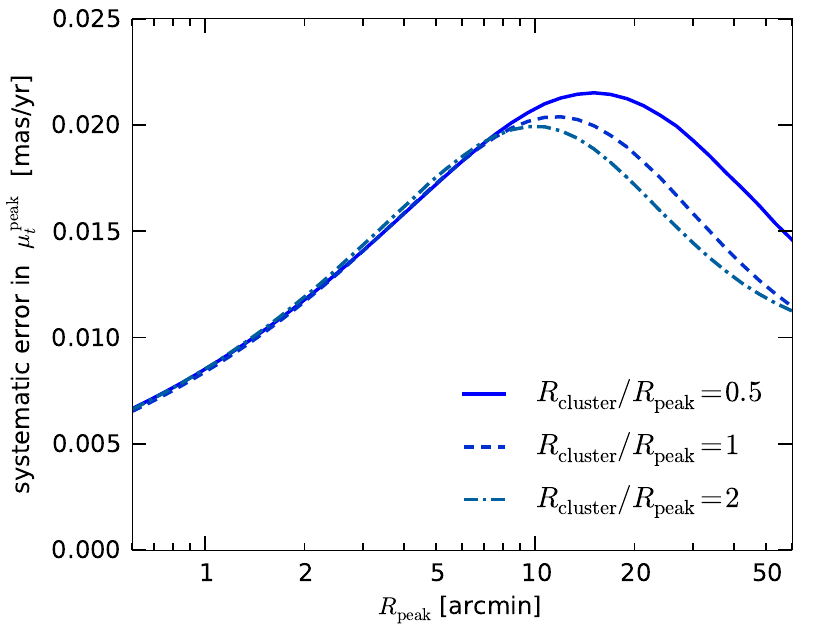}
\caption{
Systematic uncertainty on the peak amplitude of rotation, $\mu_t^\mathrm{peak}$ (Equation~\ref{eq:meanmu_parametric}) as a function of the peak radius of the rotation curve, $R_\mathrm{peak}$, for the covariance function $V_1$ (Equation~\ref{eq:covfnc}). Solid, dashed and dot-dashed lines show the results for clusters with a Gaussian surface density profile and scale radii equal to $(0.5, 1, 2)\times R_\mathrm{peak}$.
}  \label{fig:syserr_pm_rot}
\end{figure}

\begin{figure*}
\includegraphics[width=17cm]{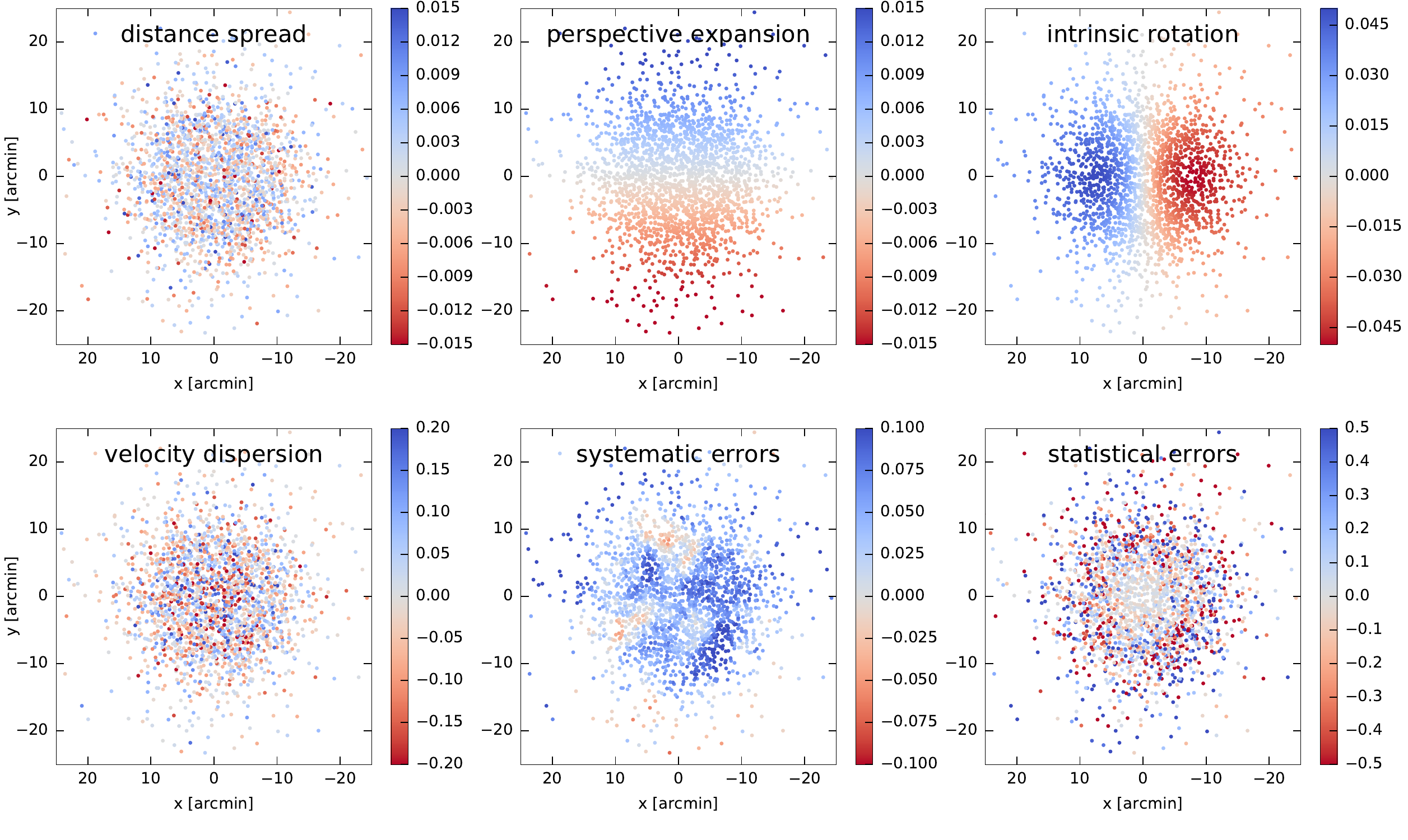}
\caption{
Example of a mock kinematic map for the globular cluster NGC~5272 (M~3).
The distance to the cluster is $10$~kpc, the line-of-sight velocity is $-150$~\kms, and the internal kinematics is described by Equations~\ref{eq:meanmu_parametric}, \ref{eq:sigma_parametric}, with $\sigma_\mu^\mathrm{centre}=0.14$~\masyr and $\mu_t^\mathrm{peak}=-0.05$~\masyr (6 and 2 \kms correspondingly).
The positions of $\sim4000$ stars and the statistical uncertainties on their PM are taken from the actual cluster. \protect\\
Different panels show contribution of several factors to the measured component of PM $\mu_\delta$ for each star (in \masyr, note the different colour scales). 
\textit{Top left}: the spread in line-of-sight distances among cluster stars negligibly affects their PM. 
\textit{Top centre}: perspective expansion due to line-of-sight motion is negligible for this cluster, but can be reliably measured in some cases, in particular, for NGC~104 and NGC~5139. 
\textit{Top right}: the intrinsic rotation in the sky plane.
\textit{Bottom left}: internal velocity dispersion.
\textit{Bottom centre}: spatially correlated systematic errors not only show a mean offset from zero, but also a significant gradient across the face of the cluster, which could be comparable to that from the intrinsic rotation.
\textit{Bottom right}: statistical uncertainties are much higher than the systematic ones, but have no bias and hence can be averaged over; in the central part of the cluster our dataset contains only the bright stars, hence their PM uncertainties are smaller.
} \label{fig:mock_map}
\end{figure*}

\begin{figure*}
\includegraphics{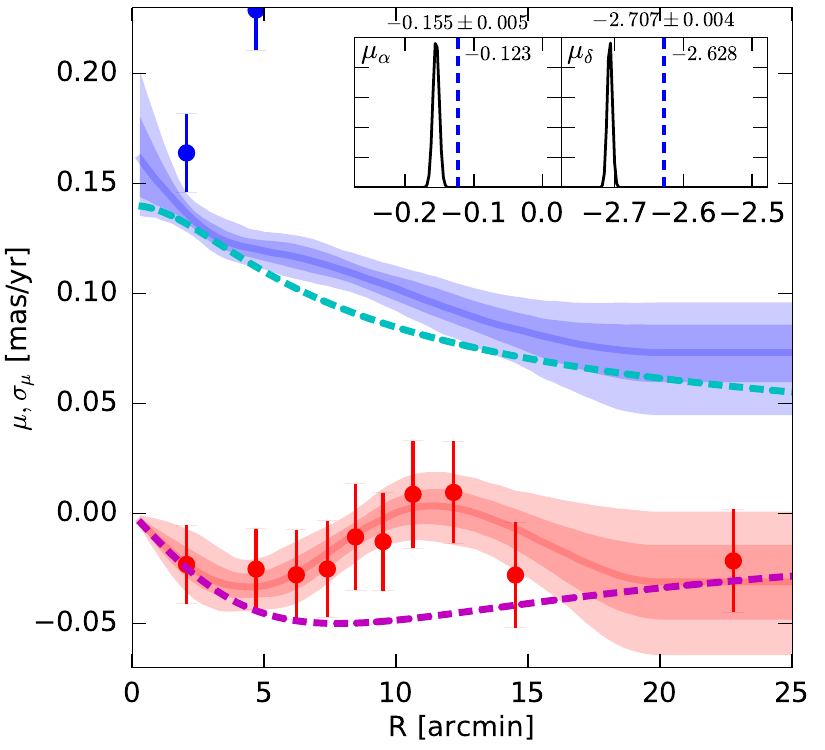}\qquad
\includegraphics{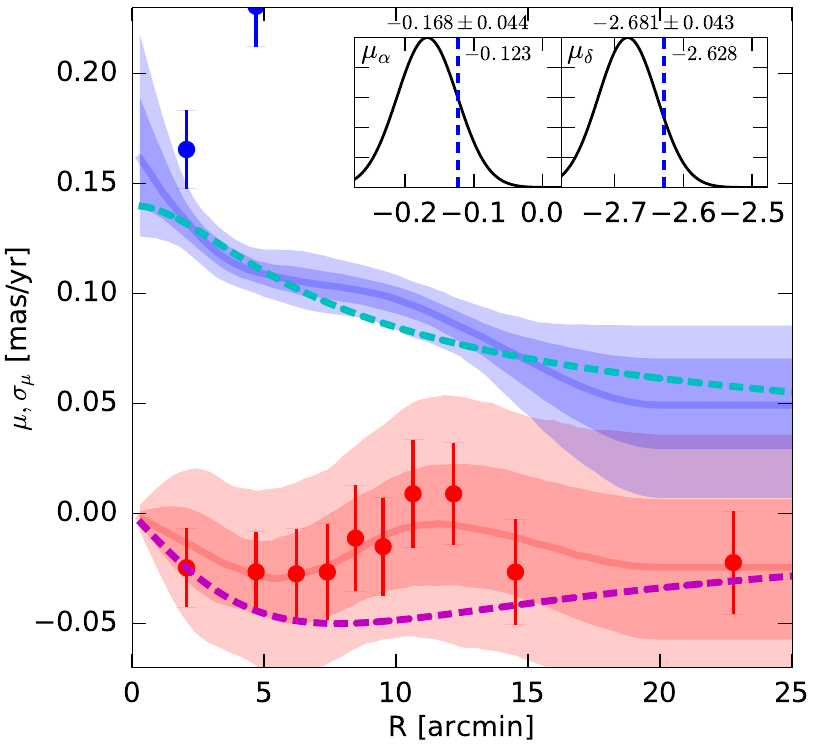}
\caption{
Example of fitting the mock kinematic map of a cluster NGC~5272 shown in Figure~\ref{fig:mock_map} with the model described in Section~\ref{sec:model}, in which both $\mu_t^\mathrm{mod}(R)$ and $\sigma_\mu^\mathrm{mod}(R)$ are represented as splines with 5 control points. \protect\\
Left panel shows the fit when ignoring the correlated systematic errors, right panel takes them into account.
Dashed magenta and cyan lines show the radial profiles of $\mu_t^\mathrm{mod}$ and $\sigma_\mu^\mathrm{mod}$ used to generate the mock data. Solid red and blue lines show the mean profiles of these quantities recovered in the fit, and corresponding darker and lighter shaded bands show the $68\%$ and $95\%$ confidence intervals. Red points show the averaged values of $\mu_t$ in 10 radial bins containing equal numbers of stars (400 per bin), with error bars reflecting the statistical uncertainties only. Blue points show the same for the raw dispersions of PM in these bins, which are due to a combination of the intrinsic dispersion and observational errors (hence is much higher than the actual $\sigma_\mu^\mathrm{mod}$ except the very centre).
The insets show the posterior probabilities of the centre-of-mass PM components $\overline\mu_\alpha$, $\overline\mu_\delta$ (in black), together with their true value (vertical dashed lines) used to construct the mock dataset. \protect\\
In this particular realization of noise, illustrated in Figure~\ref{fig:mock_map}, the systematic errors have a nonzero mean offset and a significant spatial gradient across the cluster. This nearly cancels the signal from the intrinsic rotation, as shown by the raw measurements (red dots). 
When ignoring the spatial correlations of errors in the fit, the resulting profile has too narrow confidence bands, which are determined by statistical errors only and significantly underestimate the actual deviation of the fitted profile from the true one. When these correlations are properly taken into account, the recovered profile of $\mu_t$ still does not show the rotation signature, but the confidence intervals are much wider and faithfully represent the deviation. Similarly, the centre-of-mass PM recovered in both cases are offset from the true values, but only in the latter case are consistent within the error bars. On the other hand, the recovered PM dispersion profiles are quite close to the true one in both cases, and the error bars adequately estimate the actual deviation.
} \label{fig:mock_fit} 
\end{figure*}

Before proceeding with the analysis of actual observational data, we perform extensive tests on simulated data. To start with, we estimate the systematic uncertainty on the amplitude of rotation ($\mu_t^\mathrm{peak}$ in Equation~\ref{eq:meanmu_parametric}) as a function of the spatial radius of the rotating population, $R_\mathrm{peak}$, ignoring all other factors (statistical errors and intrinsic velocity dispersion, and possible deviations from the assumed parametric form of the rotation profile). Figure~\ref{fig:syserr_pm_rot} demonstrates that the uncertainty lies in the range $0.01-0.02$~\masyr for typical cluster radii of a few arcmin. Of course, this is a lower limit on the actual uncertainty, especially for our free-form fits, which do not assume a particular functional form for the rotation profile. Nevertheless, as it turns out, the typical uncertainty on $\mu_t^\mathrm{mod}(R)$ at any radius is close to 0.02~\masyr for the majority of our clusters, when fitting both the simulated and actual data with a free-form rotation profile. Of course, the rotation signature is credible only if it is manifested coherently across the entire cluster and not just at a single node of the radial grid.

Next we validate the approach on more realistic mock catalogues created in the following way. We take the positions and statistical PM uncertainties from real stars of a given cluster, but assign the PM values by randomly drawing them from a Gaussian distribution with a given mean and covariance matrix specified by the model. The internal kinematics in the model is described by Equations~\ref{eq:meanmu_parametric} and \ref{eq:sigma_parametric}, with best-fit parameters derived for each cluster. The full error covariance matrix combines the actual statistical errors for cluster stars, spatially correlated systematic errors described by the function $V_1$ (\ref{eq:covfnc}), and the internal dispersion. We draw several random realizations of mock PM, as described in the Appendix~\ref{sec:mock_details}, and fit them with non-parametric models described above, in two regimes -- with or without accounting for systematic errors.

Figure~\ref{fig:mock_map} shown an example of such a mock map constructed for the globular cluster NGC~5272 (M~3) -- one of the clusters for which a rotation signature was detected by both \citet{Helmi2018} and \citet{Bianchini2018}. However, the measured amplitude of rotation is rather small ($\lesssim0.05$~\masyr), and could be entirely caused by gradients in the correlated systematic errors. For this particular realization of the noise (not at all uncommon), these gradients are of the same order as the rotation signal, but have the opposite sign, hence the rotational signature becomes hidden by the noise. Figure~\ref{fig:mock_fit} shows the result of fitting the models to these mock data, illustrating that we are not able to detect rotation in this case. More importantly, the uncertainties on the fitted profiles are adequately estimated only when taking the systematic errors into account, and enclose the true profile within 95\% confidence intervals.

By repeating these experiments for other clusters and with many realizations of noise, we conclude that the rotational signature measured in \Gaia data is only significant when it exceeds roughly  $0.05$~\masyr (i.e., three times higher than the systematic uncertainty in parametric fits described in the beginning of this section). When the number of cluster stars is smaller than $\sim10^3$, this detection limit is further increased due to the contribution of statistical errors. The PM dispersion profiles, on the other hand, are well recovered by the fitting procedure regardless of the inclusion or omission of systematic errors. This could be understood as follows: the PM distribution in any particular small region in the cluster may be shifted up or down by the systematic errors, but this does not affect the spread of PM values. However, in determining the rotational signal, we are comparing the mean PM across different regions of the cluster, which may be biased by the gradients in the systematic errors. Another reason is that the intrinsic PM dispersion is typically much higher than the amplitude of rotation, and hence is less affected by the systematics. 

Nevertheless, we consider $\sigma_\mu$ to be reliable only if it exceeds 0.1 at least within the half-mass radius of the cluster. Since the intrinsic dispersion is summed in quadrature with the statistical errors when comparing with the actual distribution of observed PM, it is quite sensitive to the correctness of the statistical uncertainties provided in the \Gaia catalogue. From the analysis of measured PM of quasars, \citet[Section 5.2]{Lindegren2018} and \citet[Section~3.2.2 and Figure~10]{Mignard2018} infer that the formal PM uncertainties are underestimated by $\sim10\%$; our own analysis supports this conclusion. Hence we multiply all statistical uncertainties quoted in the \Gaia catalogue by 1.1 before running the analysis; however, if these correction factors vary across different subsets of stars, the inferred dispersion still cannot be computed reliably. We performed fits of mock data with different correction factors in the range $1-1.2$ and found that the increase or decrease of the statistical uncertainties by 10\% shifts the inferred PM dispersion by $\sim 0.01-0.02$~\masyr correspondingly downwards or upwards. When $\sigma_\mu\ge 0.1$~\masyr, this shift is rather small compared to the value itself, but dispersions below 0.1~\masyr become increasingly more sensitive to the correctness of statistical uncertainties.

\subsection{Results for Milky Way clusters}  \label{sec:results}

\input{table.tex}
\input{figs.tex}

We apply the fitting approach to $\sim 80$ globular clusters that have $N\ge 20$ stars satisfying the selection criteria described in Section~\ref{sec:membership}. Of these, roughly a quarter do not pass further significance tests, having both rotation and dispersion below the thresholds established above, or inconsistent profiles derived from independent subsets of stars. The remaining clusters are listed in Table~\ref{tab:results}, and their rotation and dispersion profiles are shown in Figure~\ref{fig:results}.

In addition to our fits, we also plot the results from the literature, namely:
\begin{itemize}
\item Rotation profiles derived from \Gaia PM for 11 clusters by \citet{Bianchini2018}.
\item PM dispersion profiles for all clusters in our sample independently derived from \Gaia PM by \citet{Baumgardt2019}.
\item Rotation and PM dispersion profiles for 10 clusters independently determined by \citet{Jindal2019} using \Gaia.
\item PM dispersion profiles for 22 clusters measured in the HSTPROMO survey \citet{Watkins2015} using the \textit{Hubble} space telescope (\textit{HST}).
\item Line-of-sight velocity dispersion profiles from the MUSE IFU \citep{Kamann2018}, MIKiS survey \citep{Ferraro2018}, and a compilation of various sources and own measurements by \citet{Baumgardt2018} and \citet{Baumgardt2019}. In order to convert these data from \kms to \masyr, we used the distances determined by \citet{Baumgardt2019}, even though our own PM dispersion profiles may be somewhat different from theirs.
\end{itemize}
The first three datasets are also derived from \Gaia DR2 data, but using different selection criteria and without accounting for correlated systematic errors, hence they serve as independent consistency checks. The \textit{HST} PM are restricted to the central $1-2$ arcminutes, where the \Gaia data often suffer from crowding and incompleteness (see e.g. Figures~6, 7 in \citealt{Arenou2018}), hence they are largely complementary to our measurements.
Overall, there is a good agreement between various studies, although not without some exceptions. 

We confirm the detection of rotation in 11 clusters from the analysis of \Gaia PM by \citet{Bianchini2018}. Seven of these clusters have unambiguous rotation signatures: NGC~104 (47~Tuc), NGC~5139 ($\omega$~Cen), NGC~5904 (M~5), NGC~6273 (M~19), NGC~6656 (M~22), NGC~7078 (M~15), NGC~7089 (M~2), while the remaining four -- NGC~4372, NGC~5272 (M~3), NGC~6752, NGC~6809 (M~55) -- have weaker signatures that may be consistent with zero at 2$\sigma$ level, given the systematic errors. \citet{Jindal2019} also measured rotation in six of these clusters from \Gaia PM. We also clearly detect rotation in NGC~6266 (M~62), which was mentioned in \citet{Bianchini2018} as a probable candidate. A few other clusters also have possible signs of rotation: Pal~7 (IC~1276), NGC~5986, NGC~6093 (M~80), NGC~6341 (M~92), NGC~6388, NGC~6402 (M~14); additional tests with different subsets of stars confirm the persistence of these signatures, but their significance is low ($\lesssim 2\sigma$ level), especially for the last three clusters.
Most of these clusters have been found to be rotating based on line-of-sight velocity measurements, or their combination with \Gaia PM: NGC~4372 \citep{Kacharov2014}, NGC~5272 (\citealt{Fabricius2014}, \citealt{Ferraro2018}, \citealt{Sollima2019}), NGC~5986 \citep{Lanzoni2018}, NGC~6093, NGC~6341 (\citealt{Fabricius2014}, \citealt{Sollima2019}), NGC~6388 \citep{Lanzoni2013,Kamann2018}. 
NGC~6402 was also mentioned in \citet{Bianchini2018} as a probable candidate of insufficient significance. We could not detect any appreciable rotation in the other candidate clusters listed in that paper.

\begin{figure*}
\includegraphics{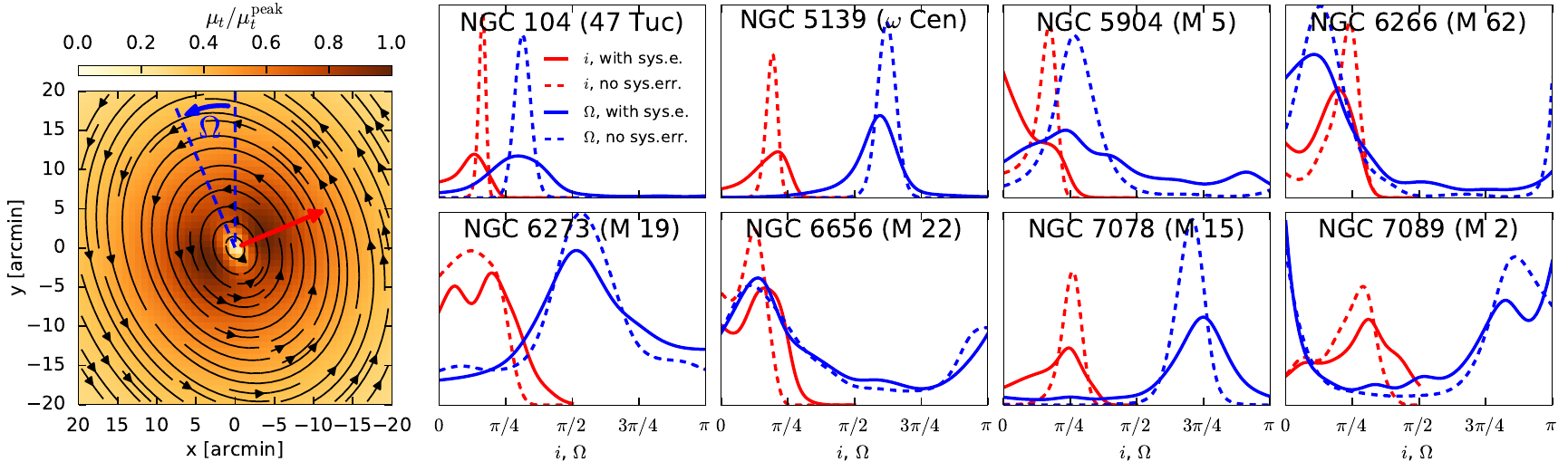}
\caption{
Constraints on the orientation of clusters with strong rotation signatures. \protect\\
Left panel illustrates the geometry of our simplified model for the rotation pattern. We assume that the stars are confined to a disc with the radial dependence of rotation velocity described by Equation~\ref{eq:meanmu_parametric}. This disc is inclined w.r.t.\ the image plane by an angle $i$ (0 means face-on orientation and $90^\circ$ -- edge-on), and the line of nodes (the intersection of the disc and image planes) is rotated by angle $\Omega$ counterclockwise from the north direction (shown in blue); projection of angular momentum vector onto the image plane (red arrow) is perpendicular to the line of nodes. The streamlines of rotational motion are shown by black arrows, and the intensity of shading shows the amplitude of the tangential component of PM $\mu_t$ on the sky plane. \protect\\
Right panels show the posterior distributions of the two orientation angles $i$ (red) and $\Omega$ (blue), derived for eight clusters with significant detected rotation. Dashed lines show the fits which do not take into account the spatially correlated systematic errors, and solid lines show the fits with account for these correlations (producing larger uncertainties in the best-fit parameters).
} \label{fig:incl}
\end{figure*}

\begin{figure}
\includegraphics{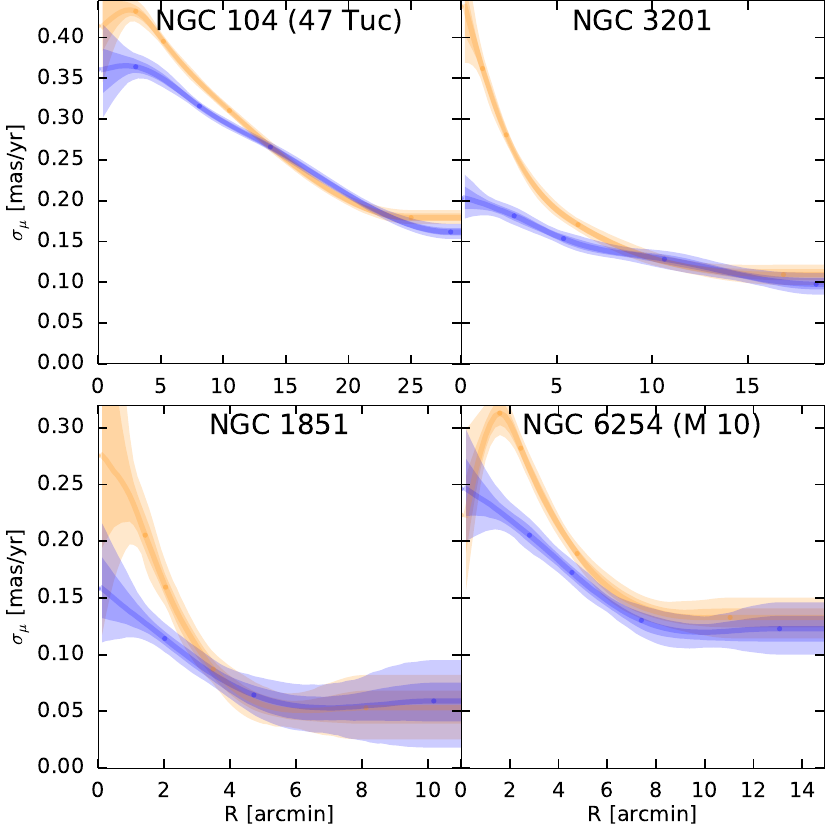}
\caption{
Comparison of PM dispersion profiles derived using different selection criteria. Top curves (orange) use stars satisfying the astrometric quality criteria (\texttt{astrometric\_excess\_noise} and renormalized unit weight error), while bottom curves (blue) additionally discard stars with high \texttt{phot\_bp\_rp\_excess\_factor} (this is the full set of quality filters used in the rest of the paper). It is clear that the stars not satisfying the last criterion (which are mostly located in crowded central regions) have underestimated uncertainties, resulting in higher inferred PM dispersions when not filtered out.
} \label{fig:sigma_compare_filters}
\end{figure}

For the eight clusters with strong rotation signatures, we attempted to constrain the orientation of the rotation axis from the asymmetries in the angular dependence of $\mu_t(R,\phi)$, where $\phi$ is the polar angle in the image plane. 
Figure~\ref{fig:incl}, left panel, illustrates the geometry of the toy model: instead of fitting the parametric profile (\ref{eq:meanmu_parametric}) to the tangential component of PM in the image plane itself, we assume that it describes the rotation profile in the equatorial plane of the cluster, which is inclined by angle $i$ to the image plane, and the intersection of the two planes (the line of nodes) is rotated by angle $\Omega$ counterclockwise from north. The projection of the intrinsic rotation axis on the image plane is perpendicular to the line of nodes. In effect, the amplitude of rotation along the line of nodes (major axis) is scaled by $\cos i$, while the amplitude of rotation along the minor axis follows the original parametric profile, but at smaller distances (also scaled by $\cos i$). When $i=0$ (face-on orientation), we recover the original parametric model with the rotation independent of $\phi$. If the stars were confined to a disc and moved on circular orbits with the prescribed radial dependence of mean velocity, this would have been a physically correct model for their observed mean PM in the image plane; since in a real cluster the stars have a complicated 3d rotation profile, we consider this model to be an ad hoc parametrization of the observed rotation signature.
Unlike \citet{vdVen2006} or \citet{Sollima2019}, we do not use the line-of-sight velocity field, only the two components of PM, hence the constraining power of this approach is not very strong.

The right panels of Figure~\ref{fig:incl} show the posterior distributions of the two orientation angles (inclination $i$ and position angle of the line of nodes $\Omega$), for the fits with or without account of systematic errors. As expected, the uncertainties are larger when considering the correlated systematic errors. For NGC~104 and NGC~5139, the preferred values of $i$ are around $30^\circ$ and $40^\circ$, correspondingly; for NGC~7089 we do not get strong constraints on the angle, while for the remaining clusters we only obtain upper limits for $i$ ranging between $45^\circ$ and $60^\circ$. Due to the simplified geometry of the model, $i$ cannot be interpreted as the inclination angle itself, but is nevertheless related to it. 
Our estimates for $i$ broadly agree with the analysis of 2d \Gaia PM maps by \citet[their Table~4]{Bianchini2018}, and with other studies: \citet{Bellini2017} determine the inclination angle of NGC~104 from the combination of line-of-sight velocity data and \textit{HST} PM as $i=30^\circ$, \citet{Watkins2013} report  $i=50^\circ$ for NGC~5039, \citet{vdBosch2006} and \citet{denBrok2014} find $i=(60\pm15)^\circ$ for NGC~7078, and \citet{Sollima2019} provide values for all eight clusters from joint fits to the line-of-sight velocity and \Gaia PM.
The position angle of the kinematic rotation axis $\Omega\pm 90^\circ$ agrees with the values determined in \citet{Bianchini2013} from the line-of-sight velocity data for NGC~104 ($136^\circ$) and NGC~5139 ($12^\circ$), but not for NGC~7078 ($106^\circ$, while our posterior peaks at $45^\circ$); however, the latter cluster may have a kinematically decoupled core \citep{vdBosch2006}, with a radial variation of position angle not captured by our simple model.

For all clusters in our sample, the radial PM component is consistent with the perspective contraction or expansion expected for the given line-of-sight velocity within the 95\% confidence interval. Only for two clusters -- NGC~3201 and NGC~5139 ($\omega$~Cen) -- the measured value is significantly different from zero (especially in the former one, which recedes at nearly 500~\kms). This agrees with the findings of \citet{Bianchini2018} concerning the radial PM component. We do not confirm the high radial gradient of $\mu_R$ in NGC~6397 found by \citet[their Figure~11]{Helmi2018}, which appears to be an artifact of an incorrect treatment of perspective effects in coordinate transformations (Bianchini et al., in prep.)

The PM dispersion in central parts of many clusters cannot be reliably measured in our data, since most of the stars are eliminated from the dataset by various quality cuts, and even the remaining ones may be affected by crowding issues. In particular, the three richest clusters in our dataset -- NGC~104 (47~Tuc), NGC~5139 ($\omega$ Cen) and NGC~6752 -- are severely affected by incompleteness in their central regions; the PM dispersion in our fits is lower than in \textit{HST} data. 
In NGC~288 our PM dispersion is higher than in \textit{HST} (although $\mu_\sigma$ hardly exceeds 0.1~\masyr even in the centre, which we consider to be a minimum credible value).
On the other hand, for the remaining 18 of clusters in common with the \textit{HST} sample, the PM dispersions agree fairly well. We consider our inferred PM dispersion for NGC~6715 (M~54) to be unreliable beyond the central two arcminutes due to difficulty in separating cluster members from stars of its parent Sgr galaxy.


Likewise, our PM dispersion profiles match quite well the binned data from \citet{Baumgardt2019} and \citet{Jindal2019}, also derived from \Gaia DR2.
In NGC~6626 (M~28) our PM dispersion is somewhat lower, while in NGC~5286, NGC~6121 (M~4) and NGC~6553 it is slightly higher than in \citet{Baumgardt2019}. Our PM dispersion profiles are lower than those of \citet{Jindal2019} for NGC~2808, NGC~5139 ($\omega$ Cen), NGC~6121 (M~4) and NGC~6752, and agree for the other six clusters considered in that paper.
M~4 is actually the closest and one of the richest ones, and its PM diagram very clearly separates its members from field stars; hence the difference in the PM dispersion between the three studies is hard to explain by underestimated statistical errors or uncertainties in membership determination. 

Comparison of PM and line-of-sight velocity dispersion profiles is the primary method for dynamical distance determination, which we did not attempt in this study, adopting the distances from the catalogue of \citet{Baumgardt2019}. The agreement between these profiles is fairly good for most clusters, with the notable exception of M~4, where our PM dispersion profile for M~4 is higher than the line-of-sight dispersion profile for the assumed distance of 2~kpc, but becomes compatible with it if the distance is smaller by some 10\% (which would better agree with most literature estimates). 

Interestingly, in many clusters we observe a flattening of PM dispersion in the outer parts, sometimes with a high level of significance (e.g., NGC~6121, NGC~6397, NGC~6752). While the PM dispersion of bound stars should drop with radius, the presence of potential escapers may inflate the measured dispersion in the outer parts \citep[e.g.,][]{Claydon2017}. Many models predict that the velocity distribution may be radially anisotropic at large radii \citep[e.g.,][]{Gieles2015}. However, measuring it unambiguously remains a serious challenge \citep{Sollima2015}, and we do not attempt this in the present study, although recently \citet{Jindal2019} found signs of radial anisotropy in the outer parts of several clusters from \Gaia PM.

We performed various consistency checks, for instance, comparing the profiles fitted for different subsets of stars (e.g., keeping only two opposite quadrants on the sky plane -- upper left and lower right, or vice versa, or selecting only the brighter half of the stars). This led to a few clusters being eliminated from the final sample, and for the remaining ones, the profiles were quite robust. We also tested the recovery of mock profiles generated from parametric fits to the data, as explained in the previous section, and found them to be satisfactory, except possibly an occasional upward bias in the central PM dispersion in a few clusters with significant crowding issues. We therefore caution against over-interpreting the steep rise in the PM dispersion at the centres of some clusters shown in our plots, although this could often be a genuine feature, e.g., in the core-collapsed cluster NGC~7078.  We also found that without applying the cut in \texttt{phot\_bp\_rp\_excess\_factor}, which eliminates many if not most stars in crowded fields, the inferred PM dispersions are often considerably higher, suggesting that the statistical uncertainties on the PM of stars with strong contamination by nearby sources may be severely underestimated (Figure~\ref{fig:sigma_compare_filters}).

\section{Discussion and conclusions}

In this work, we developed the mathematical formalism needed to take into account the spatially correlated systematic errors in \Gaia astrometry. We derived suitable approximations for the covariance function and estimated the systematic uncertainties on the mean PM and its variation within a given spatial region, as a function of the size of this region.
Using this formalism, we measured the internal kinematic profiles (PM rotation and dispersion) of several dozen Milky Way globular clusters.

Our results can be summarized as follows.
\begin{itemize}
\item The spatial covariance of systematic errors can be approximated by simple analytic expressions (Equation~\ref{eq:covfnc}, Figure~\ref{fig:covfnc}).
\item The measurement of mean PM and its gradient across a star cluster (in particular, the overall rotation) is severely affected by correlated systematic errors. Ignoring them may substantially underestimate the uncertainties on the derived parameters. 
The error-weighted estimates derived in this paper have somewhat smaller uncertainties, but are more computationally expensive than simply taking the unweighted average of pairwise covariance function as the error estimate.
Hence, a conservative rule of thumb is to add a systematic uncertainty of $0.03-0.07$~\masyr to the mean PM of a cluster, depending on its angular size (see Figure~\ref{fig:mean_pm_sys_err}), and $\sim0.02$~\masyr to the peak rotation amplitude. 
\item We detect strong rotation signatures in NGC~104, NGC~5139, NGC~5904, NGC~6266, NGC~6273, NGC~6656, NGC~7078, NGC~7089, and weaker indications of rotation in NGC~4372, NGC~5272, NGC~5986, NGC~6093, NGC~6341, NGC~6388, NGC~6402, NGC~6752, NGC~6809 and Pal~7, exceeding $0.04$~\masyr at the peak, but with less than $2\sigma$ significance.
Rotation in some of these clusters has been reported previously in \citet{Helmi2018}, \citet{Bianchini2018}, \citet{Sollima2019} and \citet{Jindal2019}.
\item For all clusters, the radial component of internal PM is consistent with perspective contraction or expansion expected for their line-of-sight velocities. For NGC~3201 and NGC~5139, the measured value is different from zero with high significance.
\item The internal PM dispersion is rather insensitive to the presence of systematic errors and to the reliability of statistical uncertainties, at least when it exceeds $\sim0.1$~\masyr. However, we find that inclusion of stars with high colour excess (an indication of contamination by the light of nearby stars) may significantly increase the inferred PM dispersion, indicating that the statistical errors may not adequately describe the scatter in the PM values for these stars (Figure~\ref{fig:sigma_compare_filters}).
\item Our PM rotation and dispersion profiles (Figure~\ref{fig:results}) generally agree with other studies based on \Gaia DR2 and \textit{HST} data \citep{Bianchini2018, Baumgardt2019, Jindal2019, Watkins2015}. Deviations mostly appear in crowded central regions, or in the cases where the PM distribution of cluster members is poorly separated from that of the field stars.
\end{itemize}

The analysis of internal PM of clusters in this work is intentionally free from any dynamical considerations, and is entirely data-driven. However, the same formalism for dealing with correlated systematic errors can be used to compute the likelihood of any physical model for the distribution function of stars, given the PM measurements of individual stars, for instance, the rotating King models of \citet{Varri2012}, anisotropic multimass King models of \citet{Gieles2015}, action-based models of \citet{Jeffreson2017}, or discrete-kinematics Jeans models of \citet{Watkins2013, denBrok2014}.

Overall, the \Gaia astrometry is a great step forward in many areas of dynamical astronomy, and the presence of systematic errors does not undermine the value of the DR2 catalogue, if properly accounted for. Hopefully, in the subsequent data releases these systematic errors will be reduced.

I thank L.~Lindegren, G.~Efstathiou, S.~Gratton and M.~Gieles for useful discussions.
This work uses the data from the European Space Agency mission \Gaia (\url{https://www.cosmos.esa.int/gaia}), processed by the \Gaia Data Processing and Analysis Consortium (\url{https://www.cosmos.esa.int/web/gaia/dpac/consortium}).
This work was supported by the European Research council under the 7th Framework programme (grant No.\ 308024). 
This work started at Aspen Center for Physics, which is supported by National Science Foundation grant PHY-1607611; my visit received support from the Simons Foundation.


\appendix

\section{Spatial correlations in fitting}  \label{sec:appendix}

This section presents the mathematical formalism for dealing with correlated systematic errors. Throughout the text, matrices are denoted by capital \textsf{Sans-serif} font, vectors -- by lowercase \textit{\textbf{boldface}}, and their elements -- as $S_{ik}$ and $b_k$.

\subsection{The likelihood function}  \label{sec:likelihood}

We assume that the joint probability distribution of the PM of all $N$ stars is a multivariate Gaussian:
\begin{equation}  \label{eq:likelihood}
\begin{array}{ll}
\mathcal{L}(\boldmu;\; \boldmu^\mathrm{mod}, \mathsf\Sigma)\!\!\!\! &\equiv
(2\pi)^{-N/2}\; (\det \mathsf\Sigma)^{-1/2} \\
&\times \exp\Big[ -\frac12 
(\boldmu - \boldmu^\mathrm{mod})^T\; \mathsf\Sigma^{-1} \; (\boldmu - \boldmu^\mathrm{mod}) \Big].
\end{array}
\end{equation}
Here $\boldmu$ are the measured PM of all $N$ stars, flattened into a single vector of length $2N$: $\{\mu_{1,x}, \mu_{1,y}, \mu_{2,x}, \dots, \mu_{N,y} \}$.
$\boldmu^\mathrm{mod}$ are the true values (or model predictions), arranged in the same way.
The covariance matrix $\mathsf\Sigma = \mathsf\Sigma_\mathrm{stat} + \mathsf\Sigma_\mathrm{sys} + \mathsf\Sigma_\mathrm{int}$ consists of three parts. The first contains $N$ two-by-two blocks along the main diagonal, with each block being the covariance matrix of statistical errors for each star:
\begin{equation}
\begin{array}{rl}
\mathsf\Sigma_\mathrm{stat}\!\!\!\! &\equiv \begin{pmatrix}
  \mathsf E_1 & \mathsf 0_{2\times2} & \cdots & \mathsf 0_{2\times2} \\
  \mathsf 0_{2\times2} & \mathsf E_2 & \cdots & \mathsf 0_{2\times2} \\
  \vdots & \vdots & \ddots & \vdots \\
  \mathsf 0_{2\times2} & \mathsf 0_{2\times2} & \cdots & \mathsf E_N
\end{pmatrix} , \\[30pt]
\mathsf E_i\!\!\!\! &\equiv \begin{pmatrix}
\epsilon_{i,x}^2 & \rho_i\, \epsilon_{i,x}\, \epsilon_{i,y} \\ 
\rho_i\, \epsilon_{i,x}\, \epsilon_{i,y} & \epsilon_{i,y}^2
\end{pmatrix},
\end{array}
\end{equation}
where $\epsilon_{i,x}, \epsilon_{i,y}$ are uncertainties on the two PM components for $i$-th star, and $\rho_i\in[-1..1]$ is the correlation coefficient quoted in the catalogue.
The second part is the covariance matrix of systematic errors:
\begin{equation*}
\begin{array}{l}
\mathsf\Sigma_\mathrm{sys} \equiv \\
\begin{pmatrix}
  V(0) & 0 & V(\theta_{12}) & \cdots & V(\theta_{1N}) & 0 \\
  0 & V(0) & 0 & \cdots & 0 & V(\theta_{1N}) \\
  V(\theta_{21}) & 0 & V(0) & \cdots & V(\theta_{2N}) & 0 \\
  0 & V(\theta_{21}) & 0 & \cdots & 0 & V(\theta_{2N}) \\
  \vdots & \vdots & \vdots & \ddots & \vdots & \vdots \\
  V(\theta_{N1}) & 0 & V(\theta_{N2}) & \cdots & V(0) & 0 \\
  0 & V(\theta_{N1}) & 0 & \cdots & 0 & V(0)
\end{pmatrix},
\end{array}
\end{equation*}
where $\theta_{ij}$ is the angular distance between $i$-th and $j$-th stars (of course, $\theta_{ij}=\theta_{ji}$), and $V(\theta)$ is the error covariance function, e.g., from Equation~\ref{eq:covfnc}. The checkerboard pattern is dictated by symmetry considerations: the systematic errors are assumed to be independent of the coordinate system orientation, hence must be the same for both PM components and uncorrelated between them.
Finally, the last part of the overall covariance matrix $\mathsf\Sigma$ is the optional internal PM dispersion: again this is a block-diagonal matrix with two-by-two blocks for each star, and in the case of isotropic but possibly spatially-varying velocity dispersion $\sigma(\boldsymbol x)$, each block is
\begin{equation}
\mathsf S_i = 
\begin{pmatrix} \sigma^2(\boldsymbol x_i) & 0 \\ 0 & \sigma^2(\boldsymbol x_i) \end{pmatrix},
\end{equation}
where $\boldsymbol x_i$ is the coordinate of $i$-th star.

In the absence of correlations between measured PM of individual stars (when $\mathsf\Sigma_\mathrm{sys}=0$), the overall covariance matrix $\mathsf\Sigma$ is a block-diagonal one, with $N$ two-by-two matrices $\mathsf\Sigma_i \equiv \mathsf E_i + \mathsf S_i$ along the diagonal. The inverse of such a matrix is also a block-diagonal matrix composed of $\mathsf\Sigma_i^{-1}$, and the determinant is $\prod_{i=1}^N \det\mathsf\Sigma_i$.
In this case, the joint likelihood function may be split into the product of $N$ separate likelihoods for each star:
\begin{equation*}
\ln \mathcal L = {-\textstyle \frac12} \sum_{i=1}^{N} \big[ \ln\det\mathsf\Sigma_i +
(\boldmu_i-\boldmu_i^\mathsf{mod})^T\; \mathsf\Sigma_i^{-1}\; (\boldmu_i-\boldmu_i^\mathsf{mod}) \big].
\end{equation*}
Naturally, this makes the computations much faster.

\subsection{The fitting procedure}  \label{sec:fitting_general}

Assume we have a particular model described by a set of parameters $\boldsymbol p$, which predicts the true PM for all stars $\boldmu^\mathrm{mod}(\boldsymbol p)$. For instance, this could be a single value for all stars, i.e., the mean PM of the cluster (of course, with two separate components, $\overline{\mu_x},\overline{\mu_y}$), or a model with some spatially-varying mean PM field $\mu_x^\mathrm{mod}(\boldsymbol x_i), \mu_y^\mathrm{mod}(\boldsymbol x_i)$, giving the prediction at each star's position $\boldsymbol x_i$. The model parameters can also include the internal PM dispersion $\sigma(\boldsymbol x_i)$.

Given the measured values $\boldmu$ and their uncertainties $\mathsf\Sigma_\mathrm{stat}, \mathsf\Sigma_\mathrm{sys}$, the fitting procedure is to find the maximum of the likelihood function (\ref{eq:likelihood}) w.r.t.\ model parameters $\boldsymbol p$.
In some special cases, the gradient of the log-likelihood w.r.t.\ model parameters can be easily expressed analytically, as in the examples below (Sections~\ref{sec:fitting_mean}, \ref{sec:fitting_example}). Then one finds the best-fit values $\boldsymbol p$ from the system of equations $\partial \ln\mathcal L / \partial\boldsymbol p=0$.
The covariance matrix of uncertainties on these parameters can be approximated as the inverse of the negative Hessian of the log-likelihood function at its maximum:
\begin{equation}  \label{eq:covar}
\mathsf C \equiv [-\mathsf H]^{-1}\,,\quad 
H_{\alpha\beta} \equiv -\frac{\partial^2 (\ln\mathcal L)}{\partial p_\alpha \;\partial p_\beta} .
\end{equation}

In other cases, it is more straightforward to maximize $\ln\mathcal L(\boldsymbol p)$ directly, using a generic multidimensional algorithm such as Nelder--Mead (also known as \textsc{amoeba}, \citealt{NumRec}, Chapter 10.5). To derive the confidence intervals on $\boldsymbol p$, one may use the MCMC approach (we employ the code \textsc{emcee} by \citealt{ForemanMackey2013}).

\subsection{The case of fixed covariance matrix}  \label{sec:fitting_mean}

Consider the task of determining the mean PM of a cluster, when the matrix $\mathsf\Sigma$ is known and fixed. In this case, the model has two parameters $\boldsymbol{p} = \{\overline{\mu_x},\overline{\mu_y}\}$, and the prediction for the vector $\boldmu^\mathrm{mod}$ is a linear function of these parameters:
\begin{equation}  \label{eq:Pmatrix}
\boldmu^\mathrm{mod} = \mathsf{P}\; \boldsymbol{p}\;,\quad
\mathsf P \equiv \underbrace{ \begin{pmatrix}
1 & 0 & 1 & 0 & \cdots & 1 & 0 \\
0 & 1 & 0 & 1 & \cdots & 0 & 1 \end{pmatrix}^T\!\!\!}_{\displaystyle N \; 2\times2 \mbox{ blocks}}
\end{equation}

The gradient of the log-likelihood function should be zero at the best-fit solution, which means
\begin{equation}
-\frac{\partial (\ln\mathcal L)}{\partial\boldsymbol{p}} = 
\big[\mathsf P^T\, \mathsf\Sigma^{-1} \big]\; \boldmu -
\big[\mathsf P^T\, \mathsf\Sigma^{-1}\, \mathsf P\big]\; \boldsymbol{p} = 0 ,
\end{equation}
and the matrix in the second square brackets is just the inverse of the covariance matrix of model parameters $\mathsf C$ (\ref{eq:covar}). 

The standard approach for inverting the symmetric covariance matrix of data uncertainties $\mathsf\Sigma$ is the Cholesky decomposition: $\mathsf\Sigma = \mathsf{L\,L}^T$, where $\mathsf L$ is a lower triangular matrix. The cost of decomposition itself is $\mathcal O (N^3)$ operations, and then the product $\mathsf\Sigma^{-1}\, \mathsf P = \mathsf L^{-T}\,\mathsf L^{-1}\, \mathsf P$ is computed in $\mathcal O(N^2)$ operations by forward- and back-substitution. This becomes important when one needs to consider more complicated models, in which the predicted vector $\boldmu^\mathrm{mod}$ is no longer a linear function of model parameters $\boldsymbol{p}$. If these parameters do not affect the matrix $\mathsf\Sigma$, its Cholesky decomposition may be precomputed in advance, and the evaluation of the likelihood for each choice of parameters still costs $\mathcal O(N^2)$ operations for the matrix-vector multiplication $\mathsf L^{-1}\,\boldmu^\mathrm{mod}(\boldsymbol p)$. However, when the matrix $\mathsf\Sigma$ is varied during the fit (for instance, by changing the third term describing the internal PM dispersion), this is no longer feasible. Nevertheless, in some special cases a different approach is possible, which still retains an $\mathcal O(N^2)$ cost.

\subsection{The case of unknown internal dispersion}  \label{sec:fitting_disp}

Consider first the case that the internal PM dispersion $\sigma$ is the same for all stars. Then $\mathsf\Sigma(\sigma) = \mathsf\Sigma_\mathrm{stat} + \mathsf\Sigma_\mathrm{sys} + \sigma^2\,\mathsf 1_{2N\times2N}$. Instead of Cholesky, we construct an eigendecomposition of the matrix $\mathsf\Sigma(0) = \mathsf Q\, \mathsf\Lambda\, \mathsf Q^{-1}$, where $\mathsf Q$ is an orthogonal matrix (meaning that $\mathsf Q^{-1} = \mathsf Q^T$) composed of eigenvectors, and $\mathsf\Lambda \equiv \diag(\boldsymbol\lambda)$ is a diagonal matrix with eigenvalues. Then the matrix $\mathsf\Sigma(\sigma)$ for any $\sigma$ can be written as
\begin{equation}
\mathsf\Sigma(\sigma) = \mathsf Q \, (\mathsf\Lambda + \sigma^2\,\mathsf 1) \, \mathsf Q^T =
\mathsf Q \, \diag(\boldsymbol{\lambda} + \sigma^2) \, \mathsf Q^T,
\end{equation}
its determinant is
\begin{equation}
\det\mathsf\Sigma = \prod_{k=1}^{2N} (\lambda_k + \sigma^2),
\end{equation}
and the product $\mathsf\Sigma^{-1}\,\boldmu$ requires only two matrix-vector multiplications, computed in $\mathcal O(N^2)$ operations:
\begin{equation}  \label{eq:invxidotmu}
\mathsf\Sigma^{-1}\,\boldmu = \mathsf Q\; \diag\Big(\frac{1}{\boldsymbol{\lambda} + \sigma^2}\Big)\; [\mathsf Q^T\,\boldmu]
\end{equation}
(of course, the symmetric product $\boldmu^T\, \mathsf\Sigma^{-1}\, \boldmu$ only needs one matrix-vector multiplication written in square brackets).

When the internal PM dispersion is allowed to vary for each star independently, this approach cannot be used. However, if we assume a fixed spatial profile for $\sigma(\boldsymbol{x}) = a\,\sigma_0(\boldsymbol x)$ and only vary its amplitude $a$, a slight modification of the above approach still retains the favourable $\mathcal O(N^2)$ scaling. We initialize a diagonal matrix $\mathsf S$ with the values $\sigma_0^2(\boldsymbol x_i)$ for each star (two rows per star), and construct a generalized eigendecomposition $\mathsf\Sigma(0) = \mathsf S\, \mathsf Q\, \mathsf\Lambda\, \mathsf Q^{-1}$. In this case, $\mathsf Q$ is no longer an orthogonal matrix (essentially, it contains eigenvectors of the matrix $\mathsf S^{-1}\,\mathsf\Sigma(0)$, which is not symmetric), but its inverse is still computed easily as $\mathsf Q^{-1} = \mathsf Q^T\,\mathsf S$. The matrix $\mathsf\Sigma$ for an arbitrary amplitude $a$ is
\begin{equation}
\mathsf\Sigma(a) = \mathsf S\, \mathsf Q\, \diag(\boldsymbol\lambda + a^2)\, \mathsf Q^T\, \mathsf S,
\end{equation}
its determinant is
\begin{equation}
\det \mathsf\Sigma = \prod_{k=1}^{2N} (\lambda_k + a^2)\,S_{kk},
\end{equation}
and the product $\mathsf\Sigma^{-1}\,\mu$ is given by (\ref{eq:invxidotmu}) with $\sigma^2$ replaced by $a^2$.
The cost of eigendecomposition is also $\mathcal O(N^3)$, with a roughly ten times larger prefactor than for the Cholesky decomposition; thus any of these methods would be prohibitively slow for datasets of more than $\sim10^4$ stars.

\subsection{A simple example}  \label{sec:fitting_example}

We now illustrate this machinery in the following scenario: find the mean value and the internal dispersion of PM for a cluster of stars. The input data consists of coordinates $\boldsymbol{x}_i$, PM measurements $\boldmu_i$, and their statistical uncertainties $\mathsf E_i$; the systematic errors have a known spatial covariance function $V(\theta)$. We assume that the intrinsic PM distribution is a Gaussian with the mean $\overline{\boldmu}$ and a width $\sigma(\boldsymbol{x}_i) = a\,\sigma_0(\boldsymbol x)$ that has a known spatial dependence but unknown normalization $a$. The task is to estimate the three free parameters of the model ($\overline{\mu_x}$, $\overline{\mu_y}$ and $a$) and their uncertainties.

We first construct the overall covariance matrix $\mathsf\Sigma = \mathsf\Sigma_\mathrm{stat} + \mathsf\Sigma_\mathrm{sys}$ and its generalized eigendecomposition with a diagonal weight matrix containing $\sigma_0^2(\boldsymbol{x}_i)$ at each star's position.  We also note that the mean PM $\boldmu^\mathrm{mod}$ predicted by the model at each star's location is a linear function of model parameters (\ref{eq:Pmatrix}). The log-likelihood function is a quadratic function of the first two parameters ($\overline\boldmu$), but a non-linear function of $a$:
\begin{equation}  \label{eq:loglikelihood_eig}
\begin{aligned}
\ln\mathcal L &= -{\textstyle\frac12} \ln\det \mathsf\Sigma - {\textstyle\frac12}
(\boldmu - \mathsf P\, \overline\boldmu)^T\, \mathsf\Sigma^{-1}(a)\, (\boldmu - \mathsf P\, \overline\boldmu) \\
&= -{\textstyle\frac12} \sum_{k=1}^{2N} 
\bigg[ \ln(\lambda_k+a^2) + \frac{ z_k^2 }{ \lambda_k + a^2 } \bigg], \\
\boldsymbol z &\equiv \mathsf Q^T\,(\boldmu - \mathsf P\, \overline\boldmu) 
= \boldsymbol y - \mathsf R\, \overline\boldmu,
\end{aligned}
\end{equation}
where the vector $\boldsymbol y \equiv \mathsf Q^T\boldmu$ of length $2N$ and the matrix $\mathsf R\equiv \mathsf Q^T\mathsf P$ of size $2N \times 2$ may be precomputed in advance.

For a given $a$, the solution for $\overline{\boldmu}$ is obtained from a linear equation:
\begin{equation}  \label{eq:meanmu_eig}
\begin{aligned}
0 = \frac{\partial \mathcal L}{\partial \overline\boldmu} &=
\mathsf R^T \diag\Big(\frac{1}{\boldsymbol\lambda + a^2}\Big) \,
\big[ \boldsymbol y - \mathsf R\, \overline\boldmu \big] , \\
\overline\boldmu &= 
\bigg[ \mathsf R^T \diag\Big(\frac{1}{\boldsymbol\lambda + a^2}\Big)\, \mathsf R \bigg]^{-1}\;
\bigg[ \mathsf R^T \diag\Big(\frac{1}{\boldsymbol\lambda + a^2}\Big)\, \boldsymbol y \bigg] .
\end{aligned}
\end{equation}
We are left with one non-linear equation for $a$:
\begin{equation}
0 = \frac{\partial \mathcal L}{\partial a} =
\sum_{k=1}^{2N} \frac{a}{\lambda_k+a^2} \; \bigg[ \frac{z_k^2(a)}{\lambda_k+a^2} - 1 \bigg] ,
\end{equation}
where $\boldsymbol z$ depends on $a$ via (\ref{eq:loglikelihood_eig}, \ref{eq:meanmu_eig}).

After the values of $\overline\boldmu$ and $a$ that maximize the likelihood have been found, we compute the Hessian of the likelihood function, and then obtain the uncertainty covariance matrix of best-fit parameters from (\ref{eq:covar}).

A \textsc{python} script illustrating this approach is available at
\url{https://github.com/GalacticDynamics-Oxford/GaiaTools}

\subsection{A general linear model for the mean PM field}  \label{sec:fitting_details}

We may use the same approach in a more general case, when the mean PM predicted by the model is not constant but varies across the cluster, if this variation may be described as a linear function of model parameters with a fixed matrix $\mathsf P$. For instance, this is the case for our free-form rotation profile represented by a cubic spline. The value of a spline function is a linear combination of $N_R-1$ basis functions $\mu^{(n)}(R)$ with free amplitudes $b_n$: 
\begin{equation*}
\mu_t^\mathrm{mod}(R) = \sum_{n=2}^{N_R} b_n\,\mu_t^{(n)}(R).
\end{equation*}
The summation index starts from 2 because the first spline node is placed at $R=0$, and the corresponding amplitude is always zero.
We have three extra free parameters -- the components of the cluster centre-of-mass motion: $\overline{\mu_x}, \overline{\mu_y}$ and $v_\mathrm{los}$; the latter only affects the radial PM component $\mu_R(R)$ through the perspective contraction or expansion (Equation~\ref{eq:mu_radial}), since we do not use the line-of-sight velocities in the fit.

The matrix $\mathsf P$, analogous to Equation~\ref{eq:Pmatrix}, has now size $2N \times (N_R+2)$ and the following structure (it depends only on the positions of stars and the choice of spline nodes, hence is precomputed at the beginning of the fit).
Each pair of rows represents the values of $\mu_{i,x},\mu_{i,y}$ for $i$-th star, produced by each of the $N_R-1$ basis functions of the rotation profile and the three basis functions of the mean PM.
The first $N_R-1$ columns of $\mathsf P$ are
\begin{equation*}
\begin{pmatrix}
\phantom{-} \frac{y_i}{R_i} \mu_t^{(2)}(R_i) &
\phantom{-} \frac{y_i}{R_i} \mu_t^{(3)}(R_i) &  \cdots & 
\phantom{-} \frac{y_i}{R_i} \mu_t^{(N_R)}(R_i) \\
         -  \frac{x_i}{R_i} \mu_t^{(2)}(R_i) &
         -  \frac{x_i}{R_i} \mu_t^{(3)}(R_i) &  \cdots &
         -  \frac{x_i}{R_i} \mu_t^{(N_R)}(R_i) \\
\cdots & \cdots & \cdots & \cdots
\end{pmatrix},
\end{equation*}
and the last three columns are
\begin{equation*}
\begin{pmatrix}
1 &  0 &  \frac{x_i}{R_i} \mu_R(R_i) \\
0 &  1 &  \frac{y_i}{R_i} \mu_R(R_i) \\
\cdots & \cdots & \cdots
\end{pmatrix}.
\end{equation*}

In addition to these $N_R+2$ free parameters of the linear model $\boldsymbol p$, we have the parameters $\boldsymbol s$ describing the radial profile of the internal PM dispersion $\sigma_\mu(R)$. As explained in Section~\ref{sec:model}, we use either a parametric profile (\ref{eq:sigma_parametric}) or a free-form profile represented by a cubic spline with $N_R$ nodes. In the first case, and only if the radial scale $R_\sigma$ is kept fixed, we may use the approach outlined in Section~\ref{sec:fitting_disp}) that avoids the need to invert the matrix $\mathsf\Sigma(\boldsymbol s)$ for every trial value of $\boldsymbol s$ (which is just a single free parameter -- the overall amplitude of dispersion $\sigma_\mu^\mathrm{centre}$). In all other cases we resort to the more general procedure with a separate Cholesky decomposition constructed for each $\mathsf\Sigma(\boldsymbol s)$.

The log-likelihood function is a quadratic form of the parameters $\boldsymbol p$, but a non-linear function of the parameters $\boldsymbol s$. The hybrid inference algorithm thus treats them differently: for each choice of $\boldsymbol s$, we determine the best-fit parameters $\boldsymbol p$ and their uncertainty covariance matrix from a linear equation, and then compute the likelihood for this vector $\boldsymbol s$. We use the MCMC approach to explore the distribution of $\boldsymbol s$ and obtain the confidence intervals on these parameters, and to compute the confidence intervals on $\boldsymbol p$, we consider all Gaussian distributions of these parameters (one for each choice of $\boldsymbol s$ taken from the Markov chain).

On the other hand, for a non-linear model such as described by the parametric rotation profile (\ref{eq:meanmu_parametric}), we treat all parameters in the same way and explore their distribution via the MCMC approach. We verified that the confidence intervals on the linear parameters $\boldsymbol p$ are nearly independent of whether we treat them separately in the hybrid procedure, or together with other parameters in the plain MCMC approach.

\subsection{The case of uncertain cluster membership}  \label{sec:membership_details}

The discussion above assumed that all observed stars belong to the cluster and hence can be described by the model for its internal kinematics. In practice, we determine the membership probabilistically, using a two-component Gaussian mixture model described in the Appendix A of \citet{Vasiliev2019}. In this model, stars are assumed to be drawn from a mixture of cluster and field populations, each one described by a Gaussian distribution $\mathcal N_c$, $\mathcal N_f$ with some unknown parameters. The prior probability of membership $q$ is also a free parameter, controlling the relative amplitudes of the two Gaussians. The attribution of each star to either of these populations is not known a priori, hence we have $N$ additional free parameters with values 0 or 1. The beauty of the Gaussian mixture approach is that one may marginalize over these unknown parameters analytically, and convert them into posterior membership probabilities $q_i$ for each star, which are not free parameters anymore, but are determined by the remaining parameters of the Gaussian distributions. The likelihood of each star is a weighted sum of probabilities of it being drawn from two alternative populations: $\mathcal L_i = q\, \mathcal N_c(\boldmu_i) + (1-q)\, \mathcal N_f(\boldmu_i)$, and the log-likelihood of the entire model is a sum of log-likelihoods of all stars (assuming that all stars are independent). The model parameters are optimized to find the maximum of the likelihood.

Unfortunately, it seems impossible to generalize this approach to the case of correlated systematic errors, because then the assumption of independence of likelihoods of each star breaks down, and one would need to marginalize over $2^N$ possible combinations of membership, each one with its own overall covariance matrix $\mathsf\Sigma$ for all stars belonging to the cluster -- this is computationally infeasible for any realistic dataset with more than 10 stars.
Therefore, we use a two-stage procedure: first determine the membership probabilities for all stars ignoring the correlated errors, then pick up stars with the membership probability $q_i$ larger than some threshold value (e.g., 0.9), and follow the approach described in previous sections with a full covariance matrix.

For most clusters, the distribution of posterior probabilities $q_i$ is very strongly bimodal, with values either close to 0 for the foreground stars or to 1 for the cluster stars, hence the exact value of the threshold does not matter. However, for $\sim20\%$ of clusters whose mean PM is close to that of the field population, the mixture model produces a broader distribution of probabilities between 0 and 1. This becomes a problem for the estimate of PM dispersion, because the genuine member stars which lie in the tails of PM distribution (where they overlap with the field population) have a lower probability of membership, and their exclusion may bias the PM dispersion downward.
We therefore adopted the following strategy: use the general model described in Section~\ref{sec:fitting_details} for the internal PM distribution of cluster stars as part of the two-component mixture model, but ignoring the correlations between stars. Then the parameters of the cluster PM distribution seem to be much less affected by this bias, because even stars with less-than-certain membership probability still contribute to the inference on the internal PM dispersion. We run the MCMC and collect the distribution of parameters $\boldsymbol s$ describing the PM dispersion profile of cluster stars $\sigma_\mu^\mathrm{mod}(R)$ from the Markov chain.
Then for each realization of these parameters $\boldsymbol s$, we pick up stars with membership probability $q_i>0.9$, construct their overall covariance matrix which now includes the systematic errors, and compute the best-fit linear parameters $\boldsymbol p$ and their confidence intervals, as described in Section~\ref{sec:fitting_details}. These confidence intervals are then averaged over the ensemble of $\boldsymbol s$ from the chain.

In effect, this compromise strategy evaluates the PM dispersion profiles while ignoring the systematic errors, but estimates the uncertainties on the rotation profiles and the centre-of-mass PM taking into account these correlated errors. We verified that the neglect of systematic errors has little impact on the PM dispersion, at least for the majority of clusters where the membership probability is strongly bimodal and hence the list of member stars is well-known. For these clusters, we ran the approach described in Section~\ref{sec:fitting_details} both with and without accounting for correlated systematic errors, and found that the uncertainties on the PM dispersion profiles were nearly identical. Of course, the MCMC simulations are much faster when we ignore correlations, because then at each step we only need to invert $N$ two-by-two matrices instead of one $2N\times2N$ matrix (recall that the cost of Cholesky decomposition scales as $N^3$).

\subsection{Generation of mock datasets}  \label{sec:mock_details}

Finally, we describe the method for generating the mock datasets used in Section~\ref{sec:mock}.
We assume a particular choice for the model parameters -- the centre-of-mass PM, radial profiles of $\mu_t^\mathrm{mod}(R)$ and $\mu_R^\mathrm{mod}(R)$ (the latter being induced by perspective effects), and the internal PM dispersion $\sigma_\mu^\mathrm{mod}(R)$.
We pick the positions of the stars and the statistical uncertainties on their PM from the actual catalogue; one may equally well assign plausible mock values for these quantities.
The overall covariance matrix $\mathsf\Sigma$ combines the individual statistical errors for each star, the correlated systematic errors between all pairs of stars, and the internal PM dispersion at each star's location, as described in Section~\ref{sec:likelihood}.
We construct the vector of mean PM values predicted by the model at each star's location $\boldmu^\mathrm{mod}$, a vector of random values $\boldsymbol\xi$ drawn from the standard normal distribution (both vectors have length $2N$), and the Cholesky decomposition of the covariance matrix, representing it as $\mathsf\Sigma = \mathsf L\,\mathsf L^T$. The mock PM values are then given by $\boldmu^\mathrm{mod} + \mathsf L\,\boldsymbol\xi$.

\label{lastpage}
\end{document}

%% file: table.tex
\begin{table}
\caption{List of clusters with PM dispersion and rotation profiles reported in Figure~\ref{fig:results}.
Radius $R_{1/2}$ encloses half of the stars in our sample, and is different from (typically larger than)
the half-mass radius of all stars.
}  \label{tab:results}
\begin{tabular}{lrrrr}
Name & {}\!\!\!\!Distance & $R_{1/2}$ & $N_\mathrm{stars}$ & $\mu_t^\mathrm{peak}$ \\
& \scriptsize [kpc] & {}\!\!\scriptsize [arcmin] & & {}\!\!\scriptsize [mas\:yr$^{-1}$] \\
\hline
NGC 104 (47 Tuc) & 4.4 & 12.1 & 10000 & -0.253 \\
NGC 288 & 10.0 & 5.0 & 2545 \\
NGC 362 & 9.2 & 5.6 & 1663 \\
NGC 1851 & 11.3 & 4.7 & 783 \\
NGC 2808 & 10.2 & 6.3 & 1593 \\
NGC 3201 & 4.5 & 9.2 & 7014 \\
NGC 4372 & 5.8 & 6.2 & 2009 & -0.054 \\
NGC 4590 (M 68) & 10.1 & 5.2 & 1448 \\
NGC 4833 & 6.2 & 4.0 & 948 \\
NGC 5139 ($\omega$ Cen) & 5.2 & 14.6 & 10000 & 0.218 \\
NGC 5272 (M 3) & 9.6 & 9.0 & 3905 & -0.047 \\
NGC 5286 & 11.4 & 3.1 & 400 \\
NGC 5904 (M 5) & 7.6 & 8.1 & 4475 & 0.119 \\
NGC 5927 & 8.2 & 3.3 & 471 \\
NGC 5986 & 10.6 & 2.9 & 435 & 0.050 \\
NGC 6093 (M 80) & 8.9 & 2.8 & 353 & 0.065 \\
NGC 6121 (M 4) & 2.0 & 8.0 & 7525 \\
NGC 6139 & 9.8 & 2.4 & 267 \\
NGC 6171 (M 107) & 6.0 & 3.8 & 1206 \\
NGC 6205 (M 13) & 6.8 & 8.4 & 3936 \\
NGC 6218 (M 12) & 4.7 & 5.7 & 3250 \\
NGC 6254 (M 10) & 5.0 & 6.6 & 4693 \\
NGC 6266 (M 62) & 6.4 & 3.8 & 447 & 0.282 \\
NGC 6273 (M 19) & 8.3 & 4.1 & 775 & 0.102 \\
NGC 6293 & 9.2 & 1.8 & 141 \\
NGC 6304 & 5.8 & 2.0 & 150 \\
NGC 6325 & 7.8 & 1.4 & 137 \\
NGC 6333 (M 9) & 8.4 & 3.1 & 254 \\
NGC 6341 (M 92) & 8.4 & 6.0 & 1803 & -0.045 \\
NGC 6352 & 5.9 & 2.9 & 836 \\
NGC 6362 & 7.4 & 5.0 & 2685 \\
NGC 6366 & 3.7 & 4.8 & 2030 \\
NGC 6388 & 10.7 & 3.2 & 460 & 0.064 \\
NGC 6397 & 2.4 & 8.8 & 10000 \\
NGC 6402 (M 14) & 9.3 & 3.4 & 655 & -0.050 \\
NGC 6441 & 11.8 & 3.1 & 103 \\
NGC 6517 & 10.6 & 1.4 & 132 \\
NGC 6535 & 6.5 & 1.3 & 198 \\
NGC 6539 & 7.8 & 2.5 & 530 \\
NGC 6541 & 8.0 & 4.0 & 772 \\
NGC 6544 & 2.6 & 4.1 & 552 \\
NGC 6553 & 6.8 & 2.6 & 314 \\
NGC 6569 & 10.6 & 2.1 & 157 \\
NGC 6624 & 7.2 & 1.5 & 126 \\
NGC 6626 (M 28) & 5.4 & 2.9 & 336 \\
NGC 6656 (M 22) & 3.2 & 9.2 & 4166 & 0.148 \\
NGC 6681 (M 70) & 9.3 & 2.1 & 250 \\
NGC 6712 & 7.0 & 2.2 & 288 \\
NGC 6715 (M 54) & 24.1 & 4.1 & 751 \\
NGC 6723 & 8.3 & 3.6 & 1228 \\
NGC 6749 & 7.8 & 2.3 & 237 \\
NGC 6752 & 4.2 & 9.7 & 10000 & -0.048 \\
NGC 6760 & 8.0 & 2.4 & 282 \\
NGC 6779 (M 56) & 9.7 & 3.4 & 725 \\
NGC 6809 (M 55) & 5.3 & 6.7 & 4061 & 0.044 \\
NGC 6838 (M 71) & 4.0 & 3.5 & 1983 \\
NGC 7078 (M 15) & 10.2 & 7.1 & 2022 & 0.105 \\
NGC 7089 (M 2) & 10.5 & 5.3 & 1296 & -0.058 \\
NGC 7099 (M 30) & 8.0 & 4.7 & 1396 \\
Pal 7 (IC 1276) & 5.4 & 2.7 & 620 & 0.067 \\
\end{tabular}
\end{table}

%% file: figs.tex
\begin{figure*}
\includegraphics{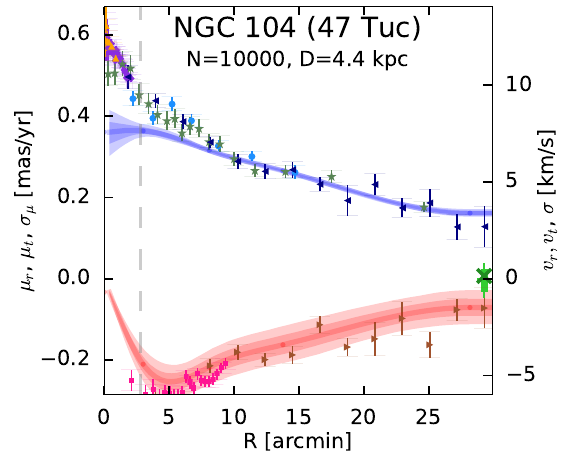}%
\includegraphics{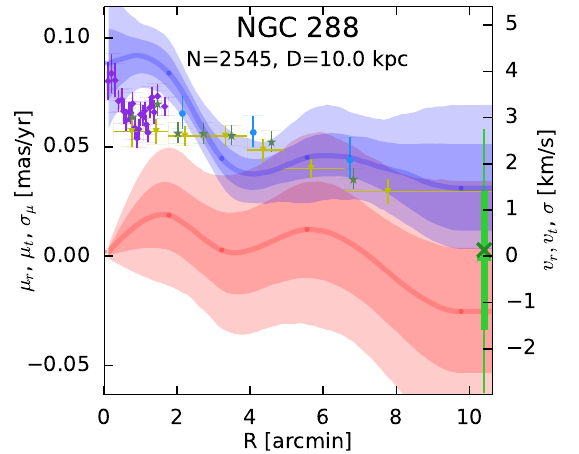}%
\includegraphics{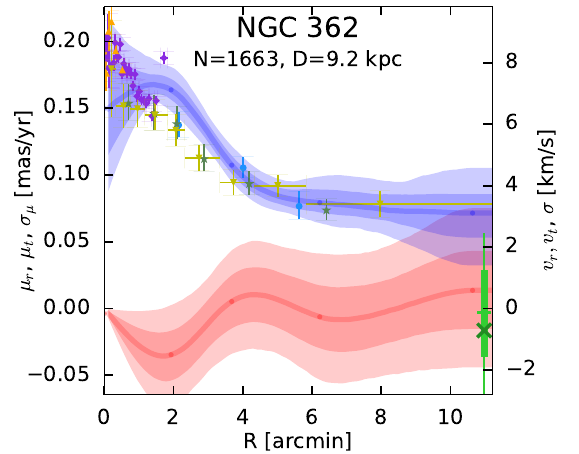}\\
\includegraphics{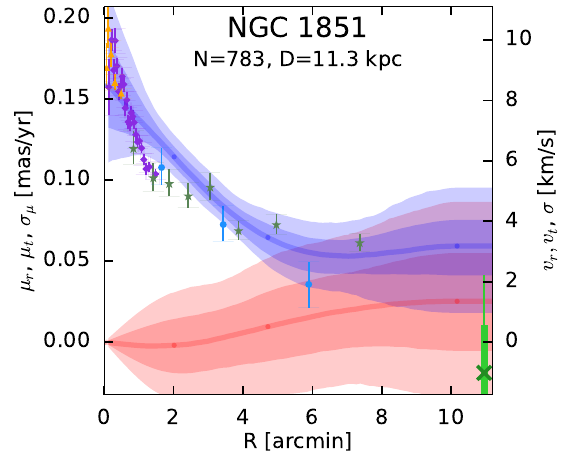}%
\includegraphics{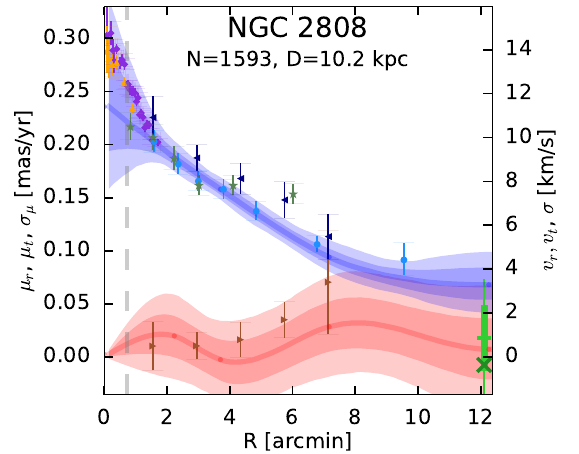}%
\includegraphics{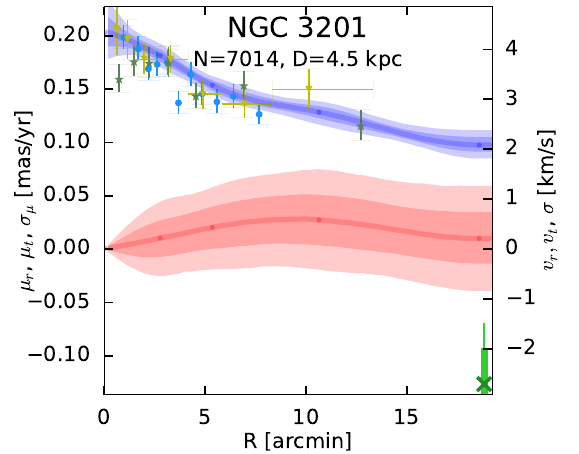}\\
\includegraphics{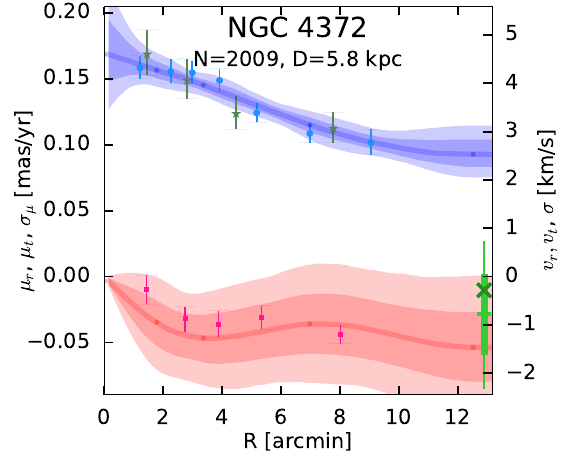}%
\includegraphics{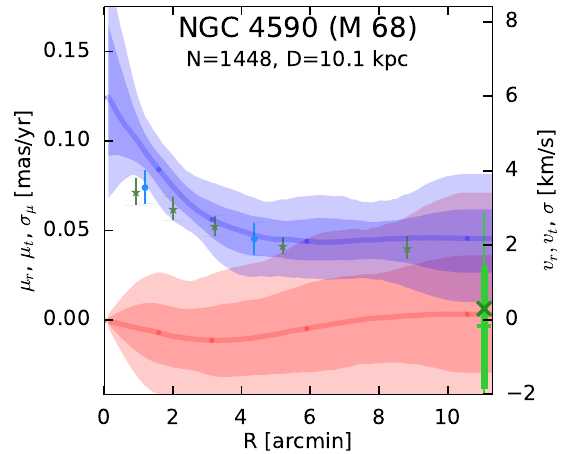}%
\includegraphics{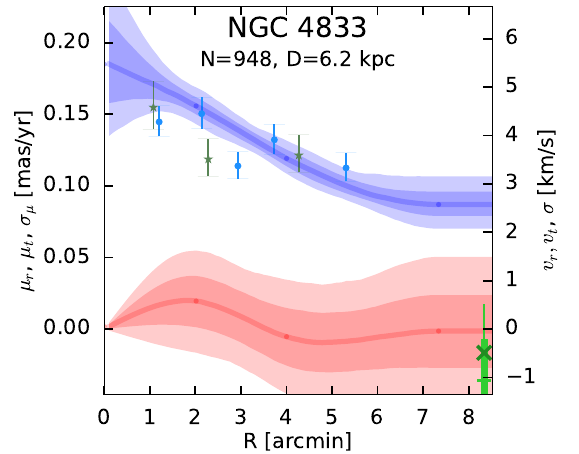}\\
\includegraphics{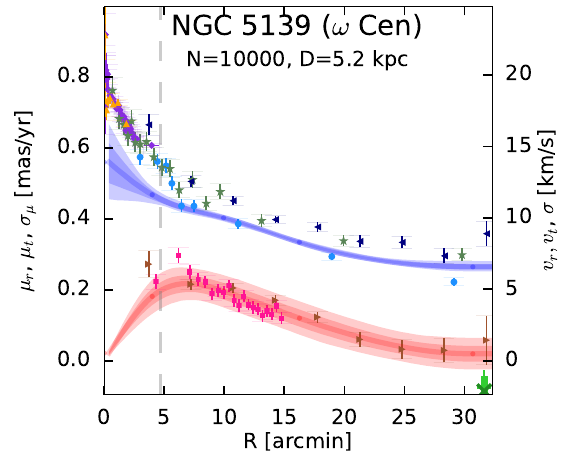}%
\includegraphics{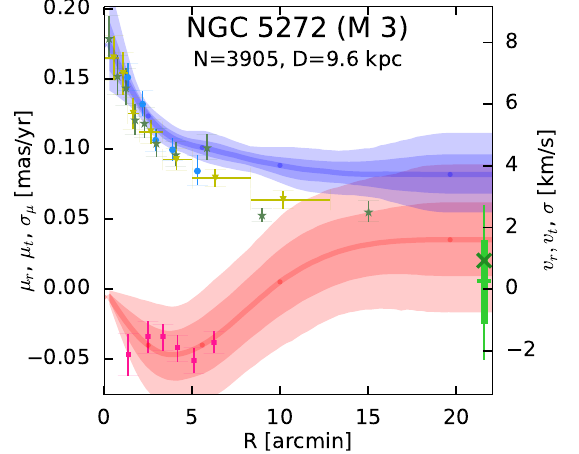}%
\includegraphics{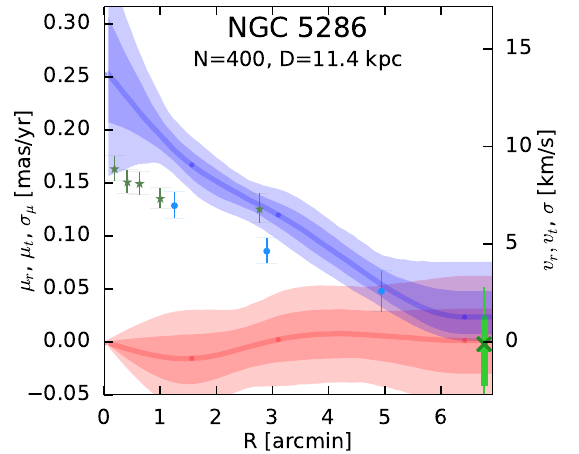}\\
\caption{
Kinematic profiles of Milky Way globular clusters derived in this work and in other studies.\protect\\
Blue and red solid lines show the radial profiles of internal PM dispersion $\sigma_\mu^\mathrm{mod}$
and mean rotation $\mu_t^\mathrm{mod}$; darker and lighter shaded bands depict 68\% and 95\% confidence
intervals taking into account systematic errors. Green error bar  shows the mean radial component of PM
from the fit (it is assumed to vary linearly with radius, hence only the value at the right boundary is
shown): thicker and thinner error bands correspond to 68\% and 95\% confidence intervals.
Green cross at the same radius shows the value of radial PM (perspective contraction/expansion)
expected for the line-of-sight velocity given in the catalogue of \citet{Baumgardt2018}.
Pink boxes show the rotation measured from \Gaia PM by \citet{Bianchini2018};
cyan circles -- PM dispersion profiles derived from \Gaia by \citet{Baumgardt2019};
brown right-pointing triangles and dark-blue left-pointing triangles -- rotation and PM dispersion
profiles determined by \citet{Jindal2019} also from \Gaia;
violet diamonds -- PM dispersion from \textit{HST} \citep{Watkins2015};
orange upward triangles, yellow downward triangles and greenish-gray stars -- line-of-sight velocity
dispersions from \citet{Kamann2018}, \citet{Ferraro2018} and \citet{Baumgardt2018}, correspondingly.
Vertical dashed line shows the half-light radii from \citet{Baumgardt2019},
while the tidal radii determined by \citet{deBoer2019} are outside the plotted radial ranges.
\textit{(Continued on next page)}
}  \label{fig:results}
\end{figure*}

\begin{figure*}
\includegraphics{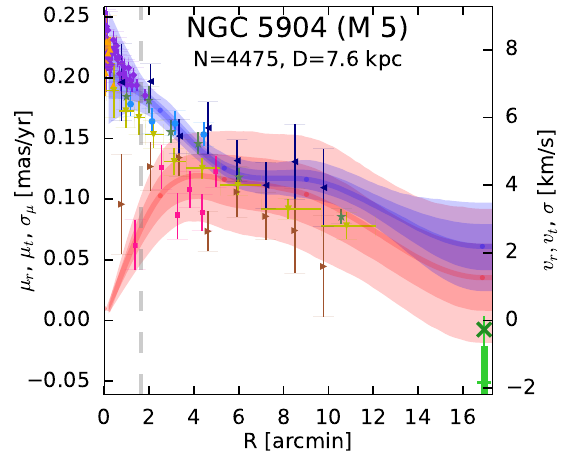}%
\includegraphics{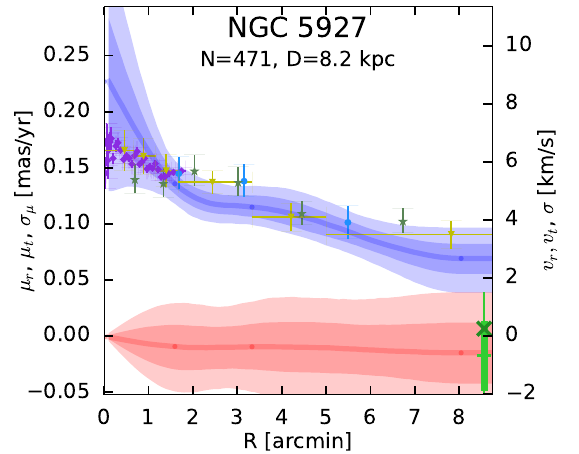}%
\includegraphics{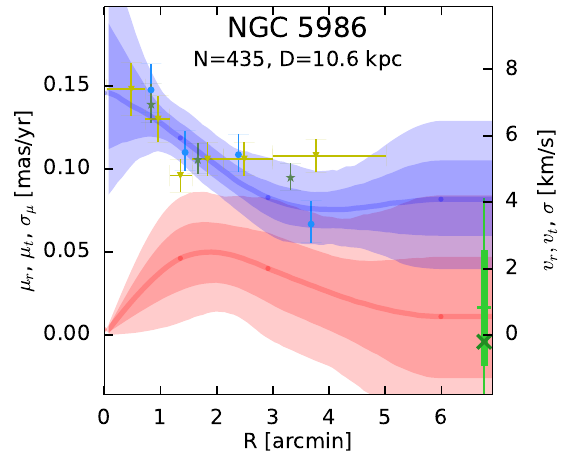}\\
\includegraphics{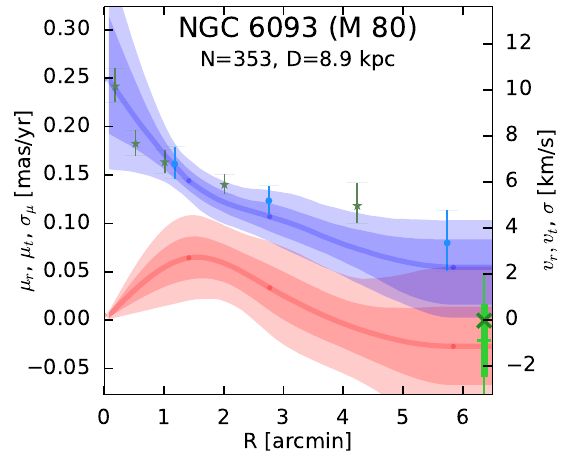}%
\includegraphics{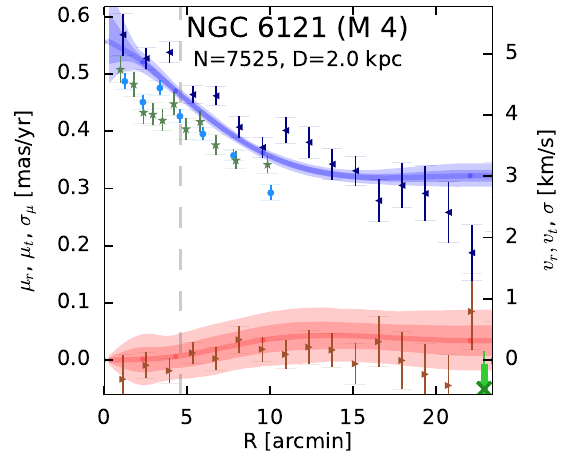}%
\includegraphics{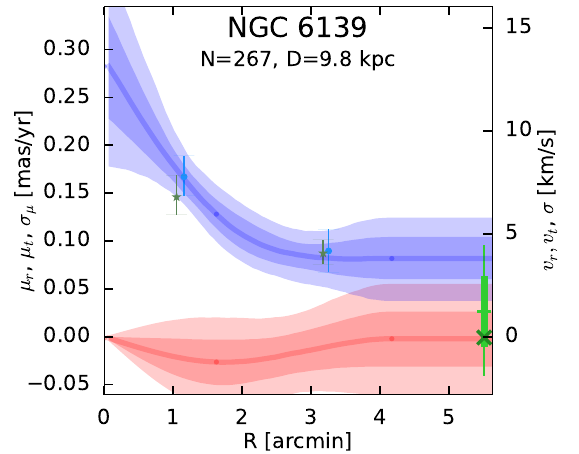}\\
\includegraphics{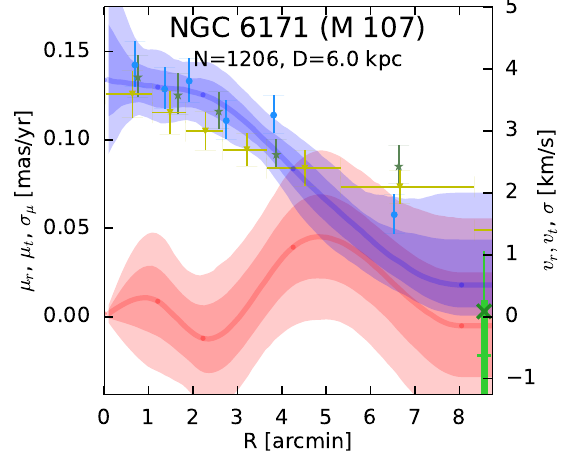}%
\includegraphics{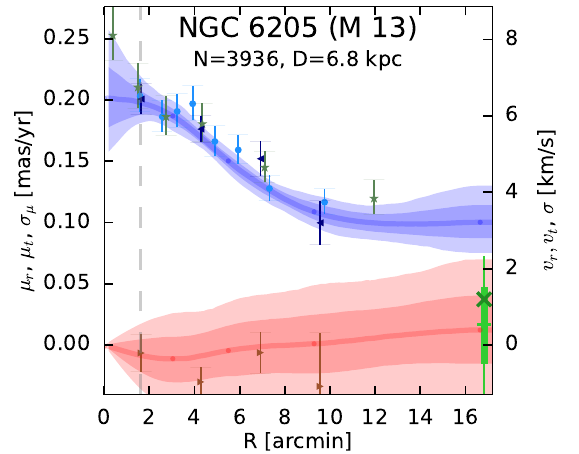}%
\includegraphics{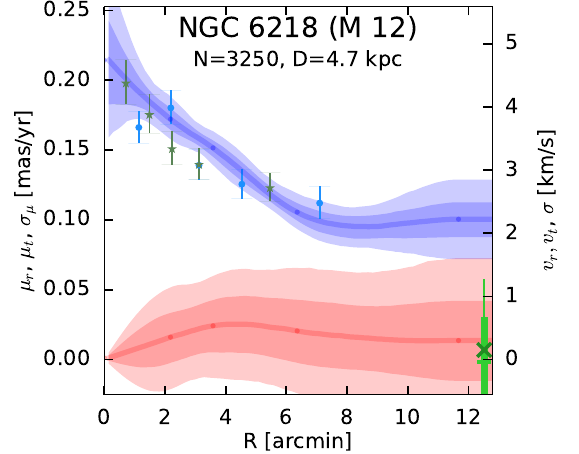}\\
\includegraphics{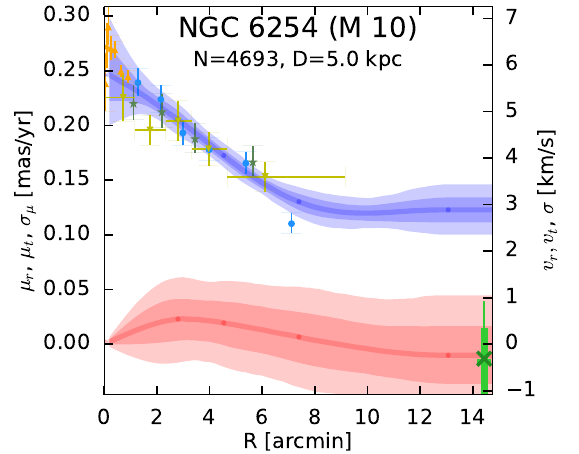}%
\includegraphics{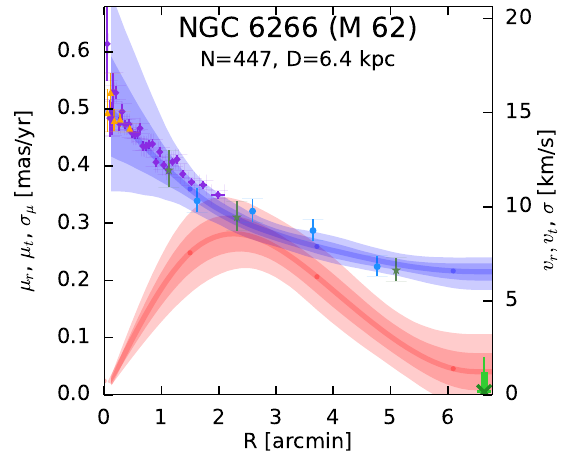}%
\includegraphics{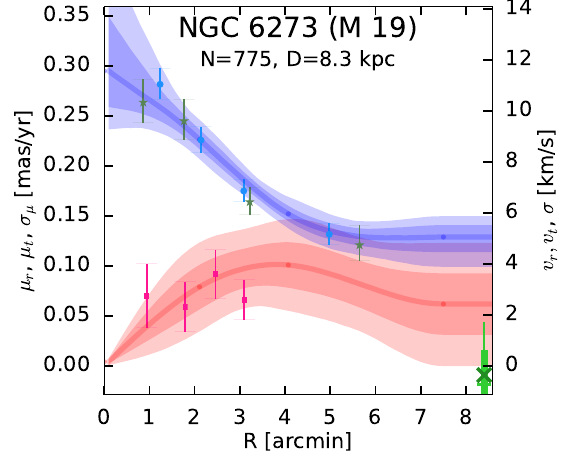}\\
\includegraphics{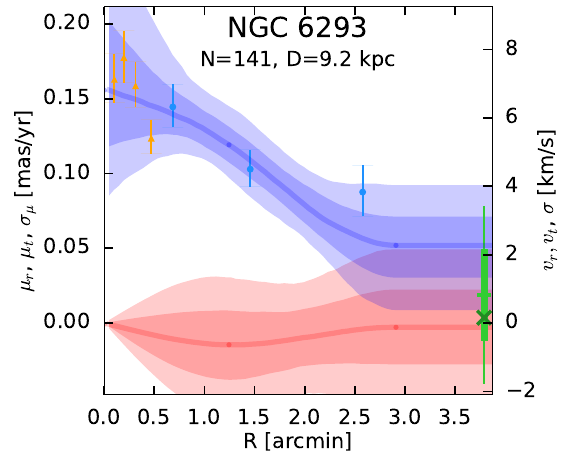}%
\includegraphics{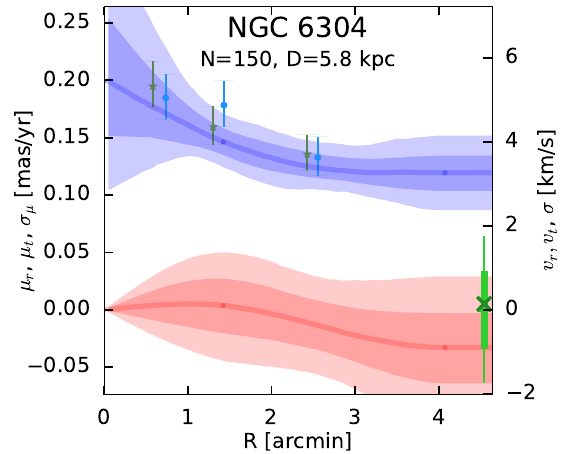}%
\includegraphics{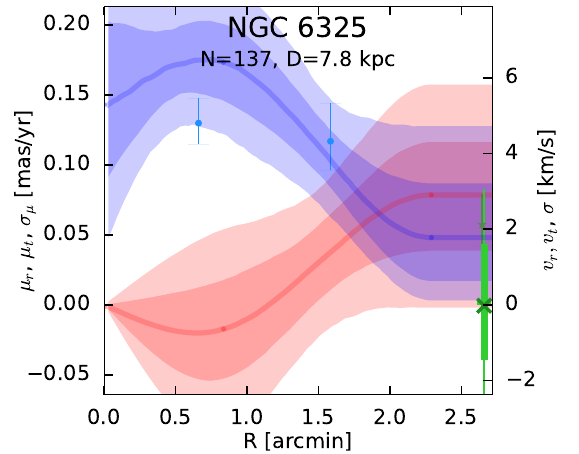}\\
\contcaption{}
\end{figure*}

\begin{figure*}
\includegraphics{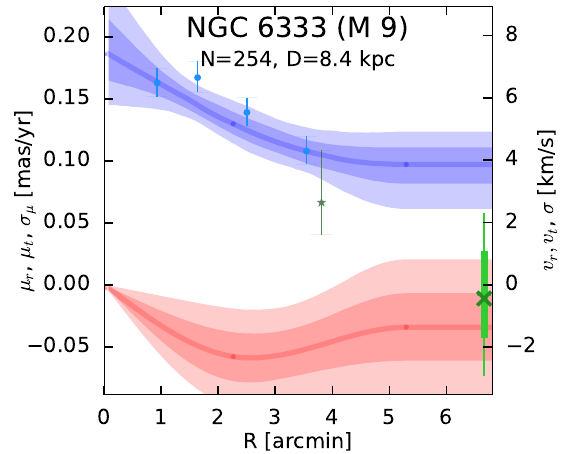}%
\includegraphics{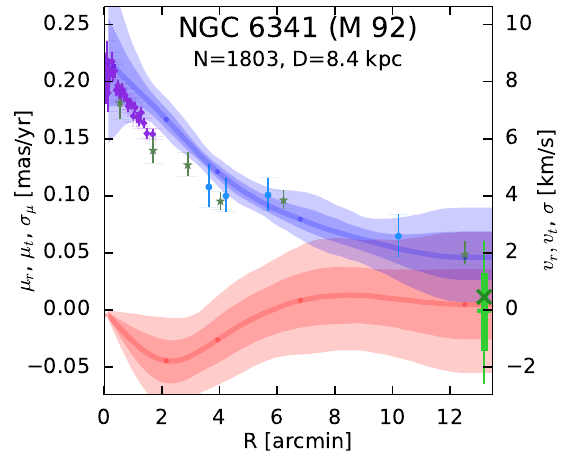}%
\includegraphics{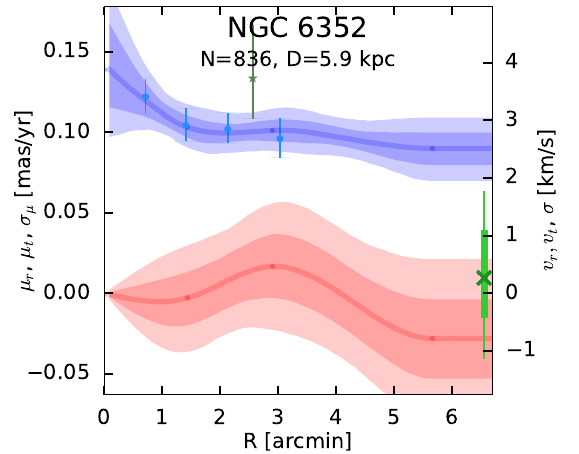}\\
\includegraphics{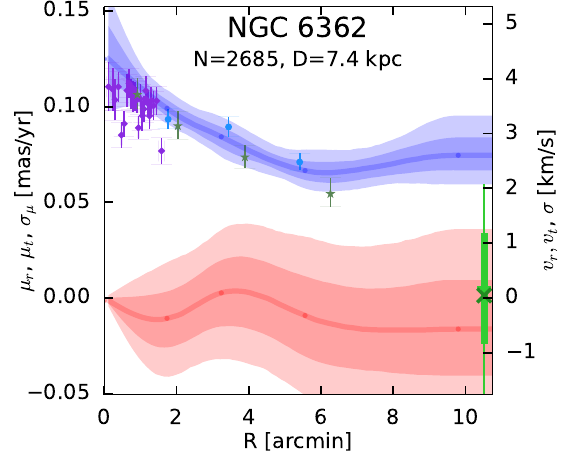}%
\includegraphics{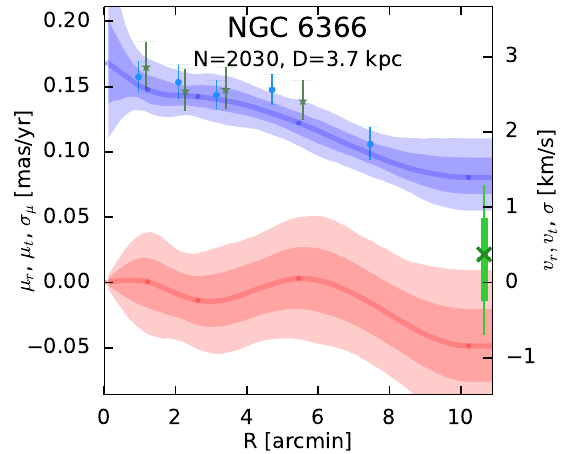}%
\includegraphics{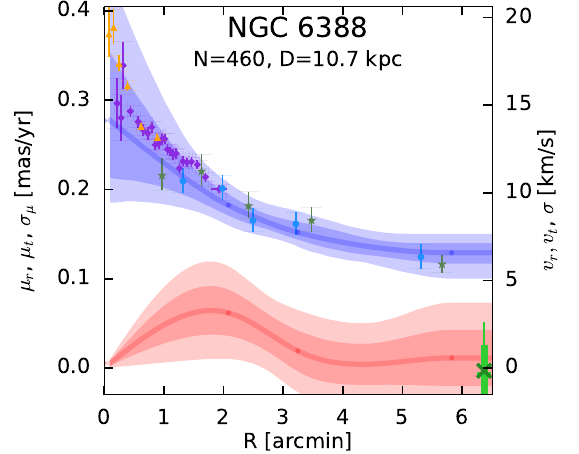}\\
\includegraphics{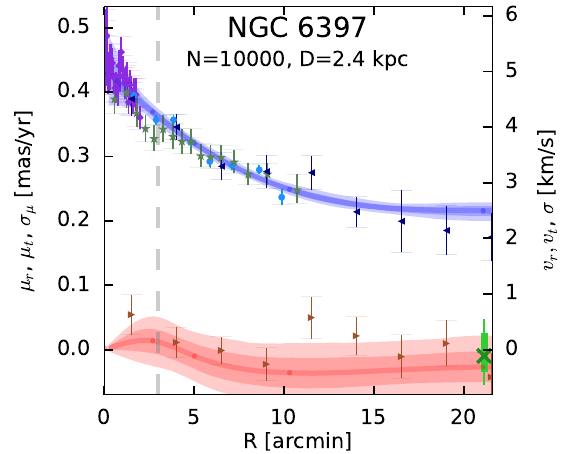}%
\includegraphics{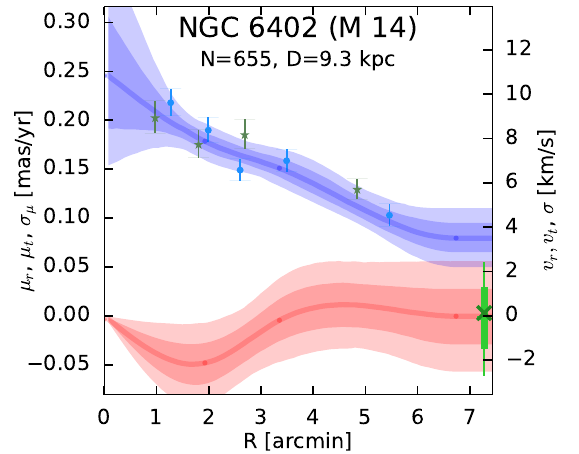}%
\includegraphics{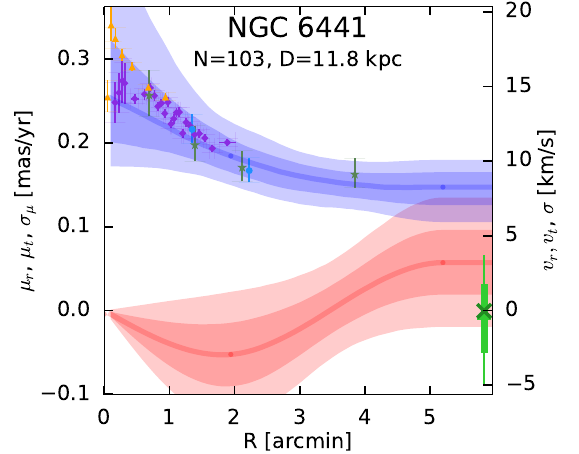}\\
\includegraphics{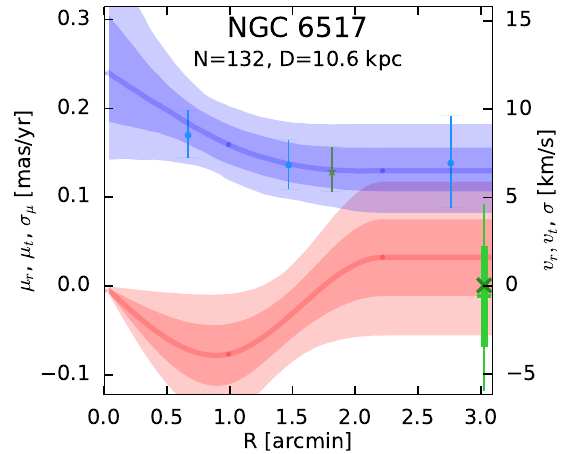}%
\includegraphics{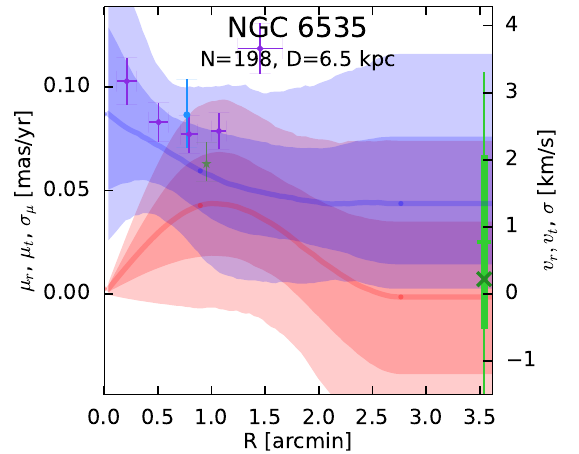}%
\includegraphics{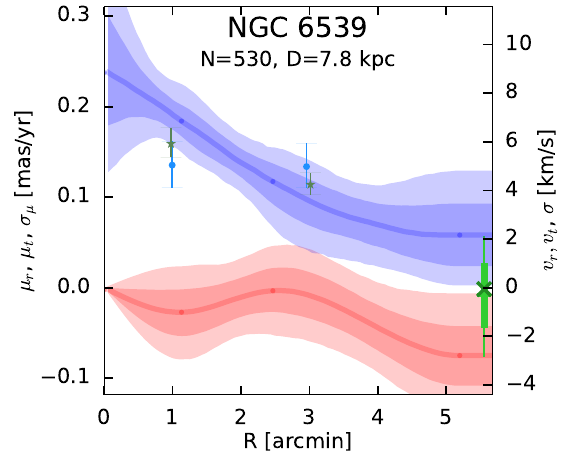}\\
\includegraphics{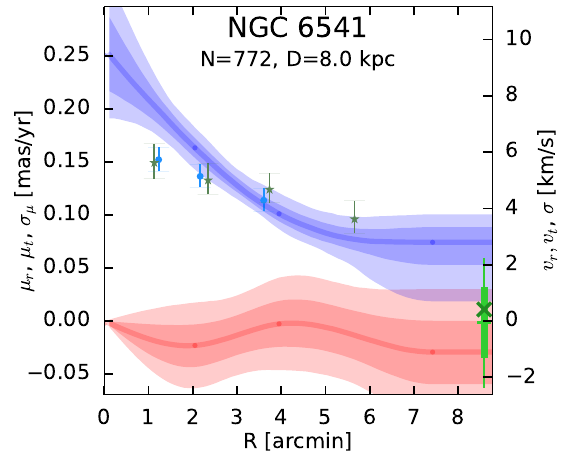}%
\includegraphics{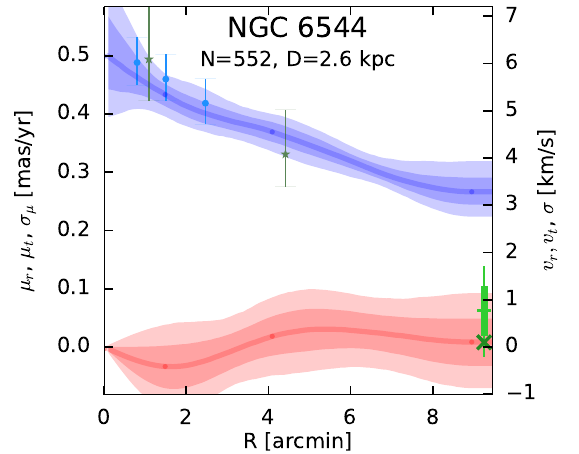}%
\includegraphics{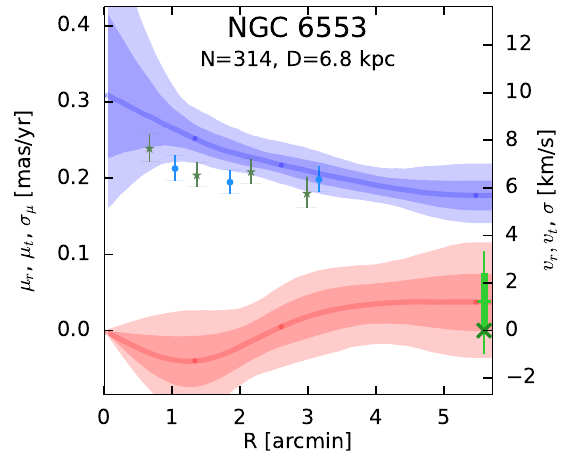}\\
\contcaption{}
\end{figure*}

\begin{figure*}
\includegraphics{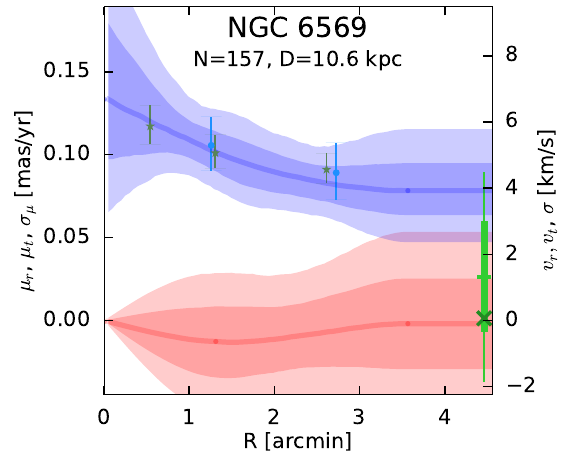}%
\includegraphics{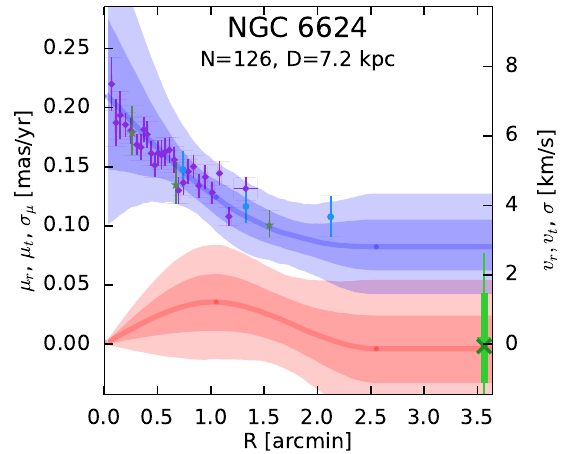}%
\includegraphics{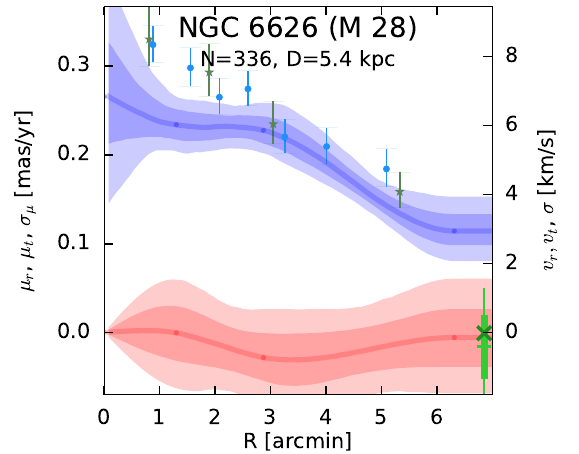}\\
\includegraphics{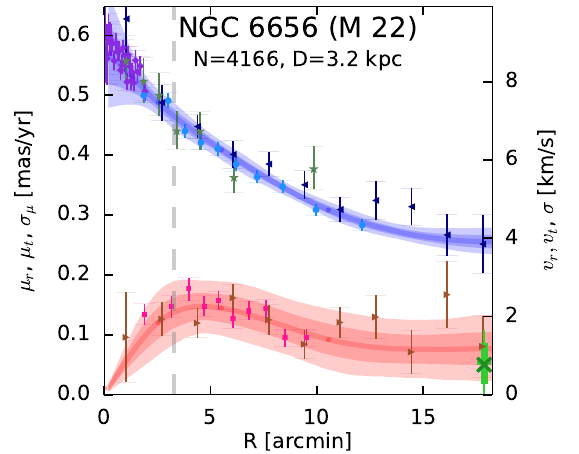}%
\includegraphics{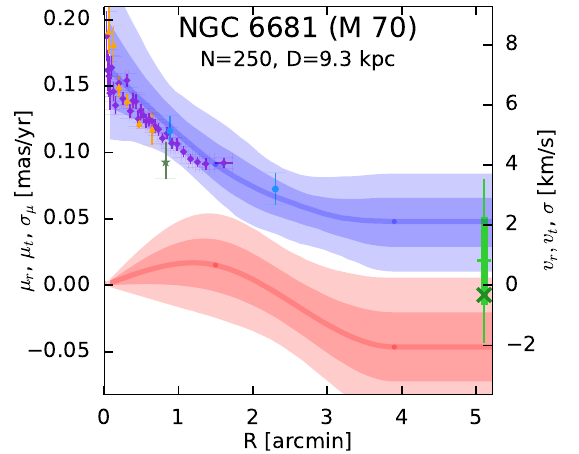}%
\includegraphics{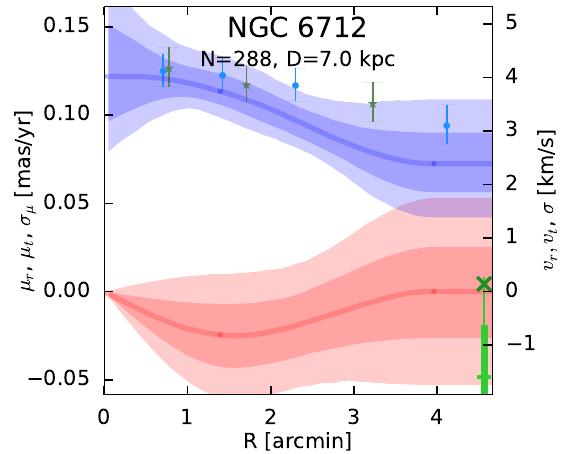}\\
\includegraphics{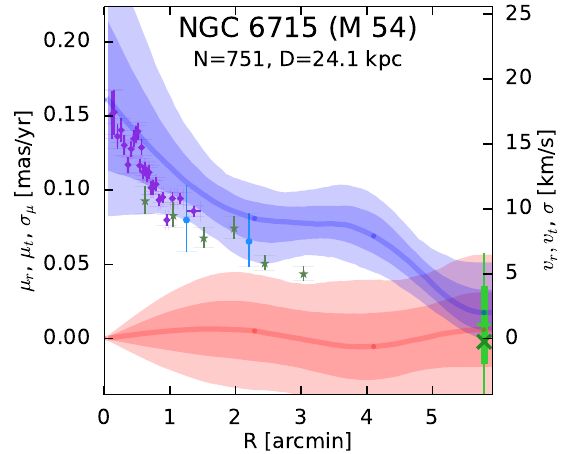}%
\includegraphics{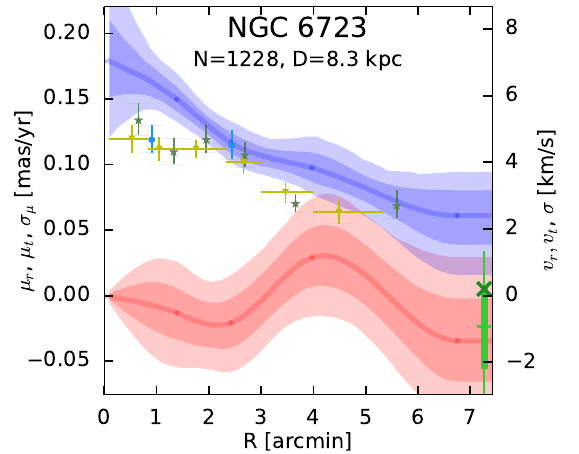}%
\includegraphics{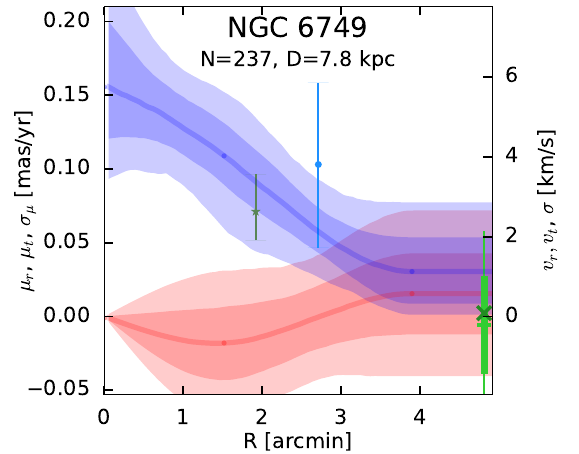}\\
\includegraphics{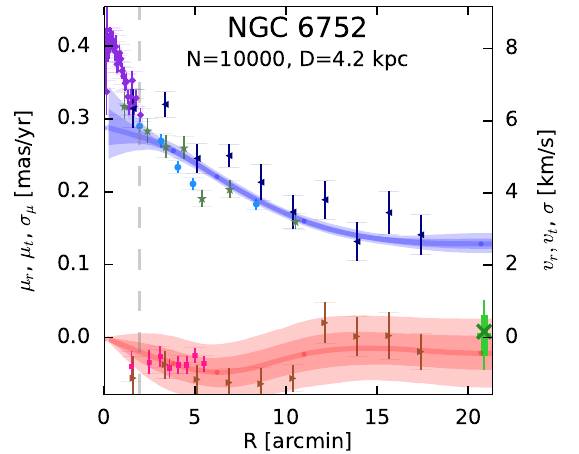}%
\includegraphics{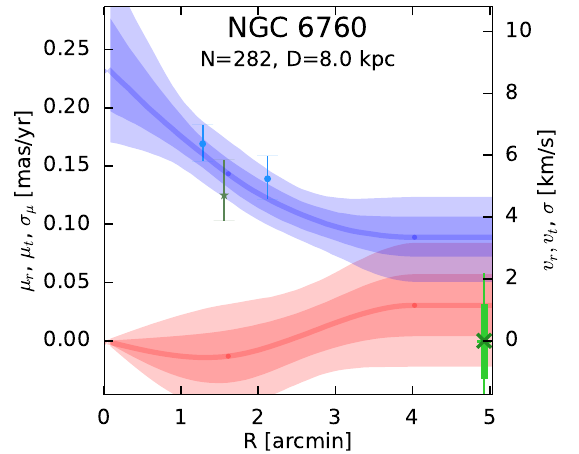}%
\includegraphics{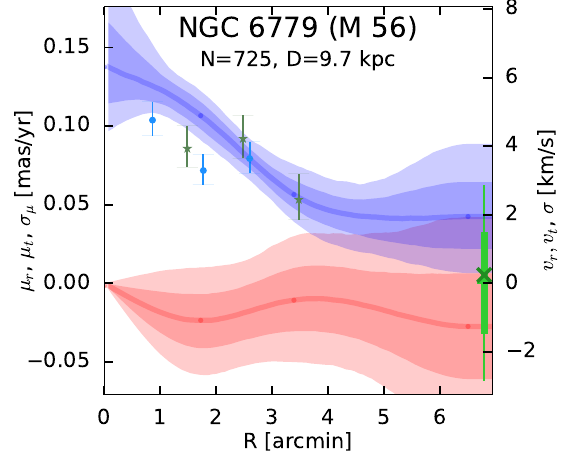}\\
\includegraphics{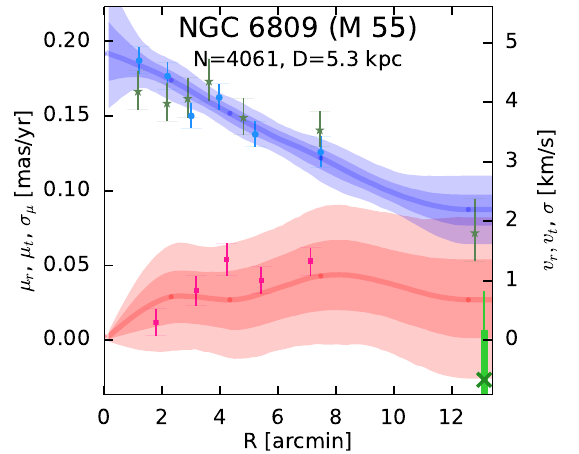}%
\includegraphics{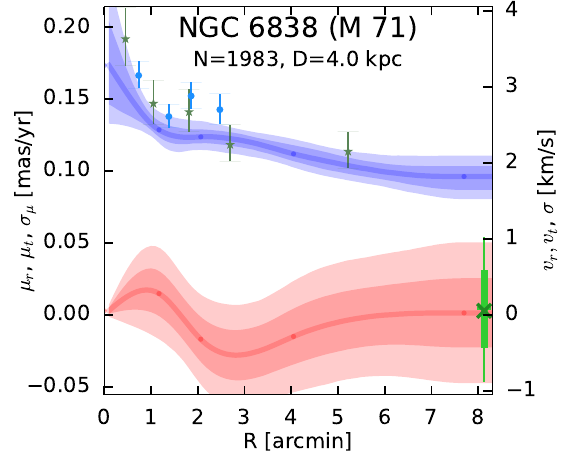}%
\includegraphics{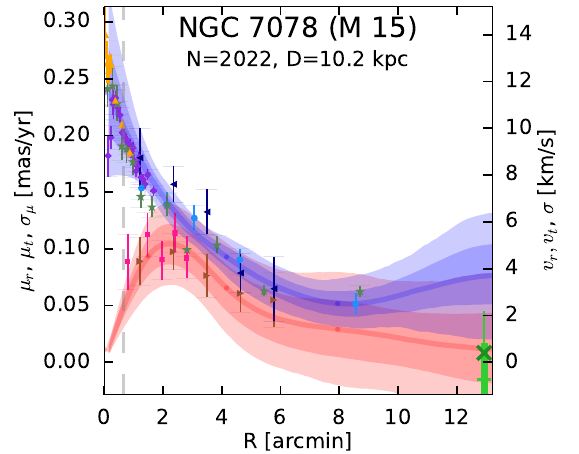}\\
\contcaption{}
\end{figure*}

\begin{figure*}
\includegraphics{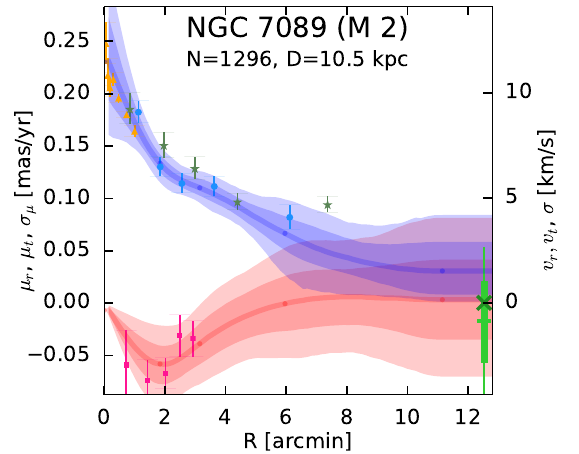}%
\includegraphics{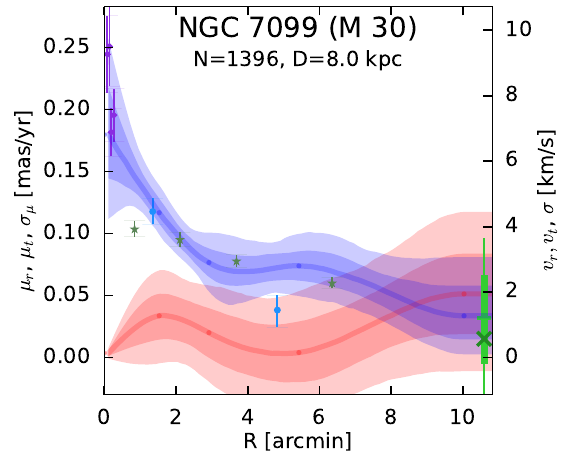}%
\includegraphics{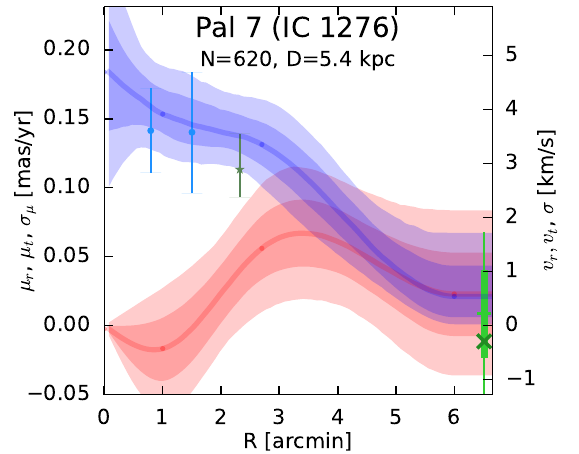}\\
\contcaption{}
\end{figure*}